%% file: main.tex
\documentclass[%
 reprint,
 twocolumn,
superscriptaddress,
showpacs,preprintnumbers,
 amsmath,amssymb,
 aps,
prd,
longbibliography
]{revtex4-1}

\usepackage{lineno}

\usepackage{tabularx}
\usepackage{multirow}
\usepackage{adjustbox}

\usepackage{graphicx}
\usepackage{subfigure}
\usepackage{dcolumn}
\usepackage{bm}
\usepackage[colorlinks = true,
            linkcolor = red,
            urlcolor  = blue,
            citecolor = red,
            anchorcolor = blue]{hyperref}

\begin{document}

\title{Measurement of the Flux-Averaged Inclusive Charged-Current Electron Neutrino and Antineutrino Cross Section on Argon using the NuMI Beam and the MicroBooNE Detector}

\newcommand{\Bern}{Universit{\"a}t Bern, Bern CH-3012, Switzerland}
\newcommand{\BNL}{Brookhaven National Laboratory (BNL), Upton, NY, 11973, USA}
\newcommand{\UCSB}{University of California, Santa Barbara, CA, 93106, USA}
\newcommand{\Cambridge}{University of Cambridge, Cambridge CB3 0HE, United Kingdom}
\newcommand{\StKates}{St. Catherine University, Saint Paul, MN 55105, USA}
\newcommand{\CIEMAT}{Centro de Investigaciones Energ\'{e}ticas, Medioambientales y Tecnol\'{o}gicas (CIEMAT), Madrid E-28040, Spain}
\newcommand{\Chicago}{University of Chicago, Chicago, IL, 60637, USA}
\newcommand{\Cincinnati}{University of Cincinnati, Cincinnati, OH, 45221, USA}
\newcommand{\CSU}{Colorado State University, Fort Collins, CO, 80523, USA}
\newcommand{\Columbia}{Columbia University, New York, NY, 10027, USA}
\newcommand{\FNAL}{Fermi National Accelerator Laboratory (FNAL), Batavia, IL 60510, USA}
\newcommand{\Granada}{Universidad de Granada, Granada E-18071, Spain}
\newcommand{\Harvard}{Harvard University, Cambridge, MA 02138, USA}
\newcommand{\IIT}{Illinois Institute of Technology (IIT), Chicago, IL 60616, USA}
\newcommand{\KSU}{Kansas State University (KSU), Manhattan, KS, 66506, USA}
\newcommand{\Lancaster}{Lancaster University, Lancaster LA1 4YW, United Kingdom}
\newcommand{\LANL}{Los Alamos National Laboratory (LANL), Los Alamos, NM, 87545, USA}
\newcommand{\Manchester}{The University of Manchester, Manchester M13 9PL, United Kingdom}
\newcommand{\MIT}{Massachusetts Institute of Technology (MIT), Cambridge, MA, 02139, USA}
\newcommand{\Michigan}{University of Michigan, Ann Arbor, MI, 48109, USA}
\newcommand{\Minnesota}{University of Minnesota, Minneapolis, MN, 55455, USA}
\newcommand{\NMSU}{New Mexico State University (NMSU), Las Cruces, NM, 88003, USA}
\newcommand{\Otterbein}{Otterbein University, Westerville, OH, 43081, USA}
\newcommand{\Oxford}{University of Oxford, Oxford OX1 3RH, United Kingdom}
\newcommand{\PNNL}{Pacific Northwest National Laboratory (PNNL), Richland, WA, 99352, USA}
\newcommand{\Pitt}{University of Pittsburgh, Pittsburgh, PA, 15260, USA}
\newcommand{\Rutgers}{Rutgers University, Piscataway, NJ, 08854, USA}
\newcommand{\StMarys}{Saint Mary's University of Minnesota, Winona, MN, 55987, USA}
\newcommand{\SLAC}{SLAC National Accelerator Laboratory, Menlo Park, CA, 94025, USA}
\newcommand{\SDSMT}{South Dakota School of Mines and Technology (SDSMT), Rapid City, SD, 57701, USA}
\newcommand{\Maine}{University of Southern Maine, Portland, ME, 04104, USA}
\newcommand{\Syracuse}{Syracuse University, Syracuse, NY, 13244, USA}
\newcommand{\TelAviv}{Tel Aviv University, Tel Aviv, Israel, 69978}
\newcommand{\Tennessee}{University of Tennessee, Knoxville, TN, 37996, USA}
\newcommand{\UTA}{University of Texas, Arlington, TX, 76019, USA}
\newcommand{\Tufts}{Tufts University, Medford, MA, 02155, USA}
\newcommand{\VTech}{Center for Neutrino Physics, Virginia Tech, Blacksburg, VA, 24061, USA}
\newcommand{\Warwick}{University of Warwick, Coventry CV4 7AL, United Kingdom}
\newcommand{\Yale}{Wright Laboratory, Department of Physics, Yale University, New Haven, CT, 06520, USA}

\affiliation{\Bern}
\affiliation{\BNL}
\affiliation{\UCSB}
\affiliation{\Cambridge}
\affiliation{\StKates}
\affiliation{\CIEMAT}
\affiliation{\Chicago}
\affiliation{\Cincinnati}
\affiliation{\CSU}
\affiliation{\Columbia}
\affiliation{\FNAL}
\affiliation{\Granada}
\affiliation{\Harvard}
\affiliation{\IIT}
\affiliation{\KSU}
\affiliation{\Lancaster}
\affiliation{\LANL}
\affiliation{\Manchester}
\affiliation{\MIT}
\affiliation{\Michigan}
\affiliation{\Minnesota}
\affiliation{\NMSU}
\affiliation{\Otterbein}
\affiliation{\Oxford}
\affiliation{\PNNL}
\affiliation{\Pitt}
\affiliation{\Rutgers}
\affiliation{\StMarys}
\affiliation{\SLAC}
\affiliation{\SDSMT}
\affiliation{\Maine}
\affiliation{\Syracuse}
\affiliation{\TelAviv}
\affiliation{\Tennessee}
\affiliation{\UTA}
\affiliation{\Tufts}
\affiliation{\VTech}
\affiliation{\Warwick}
\affiliation{\Yale}

\author{P.~Abratenko} \affiliation{\Tufts} 
\author{M.~Alrashed} \affiliation{\KSU}
\author{R.~An} \affiliation{\IIT}
\author{J.~Anthony} \affiliation{\Cambridge}
\author{J.~Asaadi} \affiliation{\UTA}
\author{A.~Ashkenazi} \affiliation{\MIT}\affiliation{\TelAviv}
\author{S.~Balasubramanian} \affiliation{\Yale}
\author{B.~Baller} \affiliation{\FNAL}
\author{C.~Barnes} \affiliation{\Michigan}
\author{G.~Barr} \affiliation{\Oxford}
% ...

\collaboration{The MicroBooNE Collaboration}
\thanks{microboone\_info@fnal.gov}\noaffiliation


\begin{abstract}
We present a measurement of the combined $\nu_e$ + $\bar{\nu}_e$ flux-averaged charged-current inclusive cross section on argon using data from the MicroBooNE liquid argon time projection chamber (LArTPC) at Fermilab. Using the off-axis flux from the NuMI beam, MicroBooNE has reconstructed 214 candidate $\nu_e$ + $\bar{\nu}_e$ interactions with an estimated exposure of 2.4$\times10^{20}$ protons on target. Given the estimated purity of 38.6\%, this implies the observation of 80 $\nu_e$ + $\bar{\nu}_e$ events in argon, the largest such sample to date. The analysis includes the first demonstration of a fully automated application of a dE/dx-based particle discrimination technique of electron and photon induced showers in a LArTPC neutrino detector. We measure the $\nu_e + \bar{\nu}_e$ flux-averaged charged-current total cross section to be ${6.84\pm\!1.51~\textrm{(stat.)}\pm\!2.33~\textrm{(sys.)}\!\times\!10^{-39}~\textrm{cm}^{2}/~\textrm{nucleon}}$, for neutrino energies above 250~MeV and an average neutrino flux energy of 905~MeV when this threshold is applied. The measurement is sensitive to neutrino events where the final state electron momentum is above 48~MeV/c, includes the entire angular phase space of the electron, and is in agreement with the theoretical predictions from \texttt{GENIE} and \texttt{NuWro}. This measurement is also the first demonstration of electron neutrino reconstruction in a surface LArTPC in the presence of cosmic ray backgrounds, which will be a crucial task for surface experiments like those that comprise the Short-Baseline Neutrino (SBN) Program at Fermilab.


\end{abstract}

\maketitle


\section{Introduction } 

The measurement of electron neutrinos ($\nu_e$) appearing in a muon-neutrino beam 
is the cornerstone of current and future accelerator-based neutrino oscillation experiments.
The appearance oscillation channel allows long-baseline experiments to determine the neutrino mass ordering~\cite{NOvA:2018gge} and to search for CP violation in the neutrino sector~\cite{dune_cp,hyperk_cpv_search}. It further allows short-baseline experiments to shed light on the possible existence of sterile neutrinos~\cite{sbnproposal}.
The success of these experiments relies on a precise understanding of electron-neutrino interactions with the detector target.
LArTPCs are being employed to perform all of the above-mentioned measurements. MicroBooNE~\cite{microboone_experiment} and ICARUS~\cite{sbnproposal} are already running, while SBND~\cite{sbnproposal} and DUNE~\cite{dune_vol2} are under construction. However, only ArgoNeuT~\cite{Acciarri:2020lhp} has made a measurement of electron-neutrino interactions on argon which included a sample of 13 selected events. In addition, only a handful of measurements of electron-neutrino interactions on other nuclei in the hundred MeV to GeV range are available \cite{gargamelle,t2k_nue,Minerva_nue}. 

The lack of precise electron-neutrino cross section measurements has been mitigated in short-baseline oscillation measurements by using $\nu_{\mu}$ interactions to constrain the oscillated $\nu_{e}$ flux and cross section models \cite{PhysRevD.79.072002,miniboone_lee}. In such an approach, any uncertainty on the $\nu_{e}$/$\nu_{\mu}$ cross section ratio would reduce the strength of the constraint that the $\nu_{\mu}$ can provide to a $\nu_{e}$ measurement. These differences, predicted to be on the order of 10\%, arise from the different final state lepton mass, radiative corrections, and modifications to the pseudo-scalar form factor \cite{Day:2012gb}. The last of these effects can be difficult to calculate. Similarly, recent theoretical calculations of the $\nu_e$ charged-current (CC) to $\nu_{\mu}$ CC cross section ratios \cite{Nikolakopoulos:2019qcr} predict differences of as much as 25\% between the $\nu_{e}$ and $\nu_{\mu}$ cross sections, particularly for forward-going leptons in the sub-GeV range. The uncertainty on the electron neutrino and antineutrino interaction model can be responsible for the majority of the uncertainty of the oscillation measurement~\cite{Abe:2019vii}. 
Independent direct measurements of electron-neutrino cross sections are therefore crucial to further inform our understanding of different flavor neutrino interactions. Performing these measurements with high precision requires suppression of backgrounds consisting of photon showers. This can be done using the amount of energy deposited per unit length at the origin of the shower, usually referred to as dE/dx, combined with the distance of the shower from the interaction vertex. Both of these methods are strengths of LArTPC detectors and their use is demonstrated in this work.    

In this paper, we present the first measurement of the electron neutrino and antineutrino charged-current cross section on argon using the MicroBooNE \cite{microboone_experiment} detector at Fermilab.
We employ neutrinos from the Neutrinos from the Main Injector (NuMI) neutrino beam \cite{numibeam} running in its ``forward horn current" mode which selects neutrinos over anti-neutrinos for the on-axis component of the flux. The measurement is performed for $\nu_e$ + $\bar{\nu}_e$ energies above 250~MeV and an average neutrino flux energy at MicroBooNE of 905~MeV with this threshold. We select an inclusive sample of interactions defined in Section~\ref{section_fv_cut} requiring at least one reconstructed electromagnetic shower inside of the fiducial volume. 

While in principle, the L/E of the NuMI beam with MicroBooNE is similar to that of the Booster Neutrino Beam (BNB), an analogous oscillation search using the NuMI beam is not practical given its significantly larger intrinsic electron neutrino content and flux uncertainties.

This paper is structured as follows: first, we briefly describe the MicroBooNE experiment (Section~\ref{sec:microboone_expt}) and the main features of the NuMI neutrino beam at the MicroBooNE detector. We then describe the simulation and reconstruction chain (Section~\ref{sec:sim_and_reco}), the event selection criteria, and their performance (Section~\ref{sec:selection}). We report the measured cross section (Section~\ref{sec:xsec_measurement}) and conclude with a discussion of systematic uncertainties (Section~\ref{sec:sys_uncertainty}).

\begin{figure}[h!]
    \center
    \hspace*{-2cm}
    \includegraphics[width=1.37\linewidth]{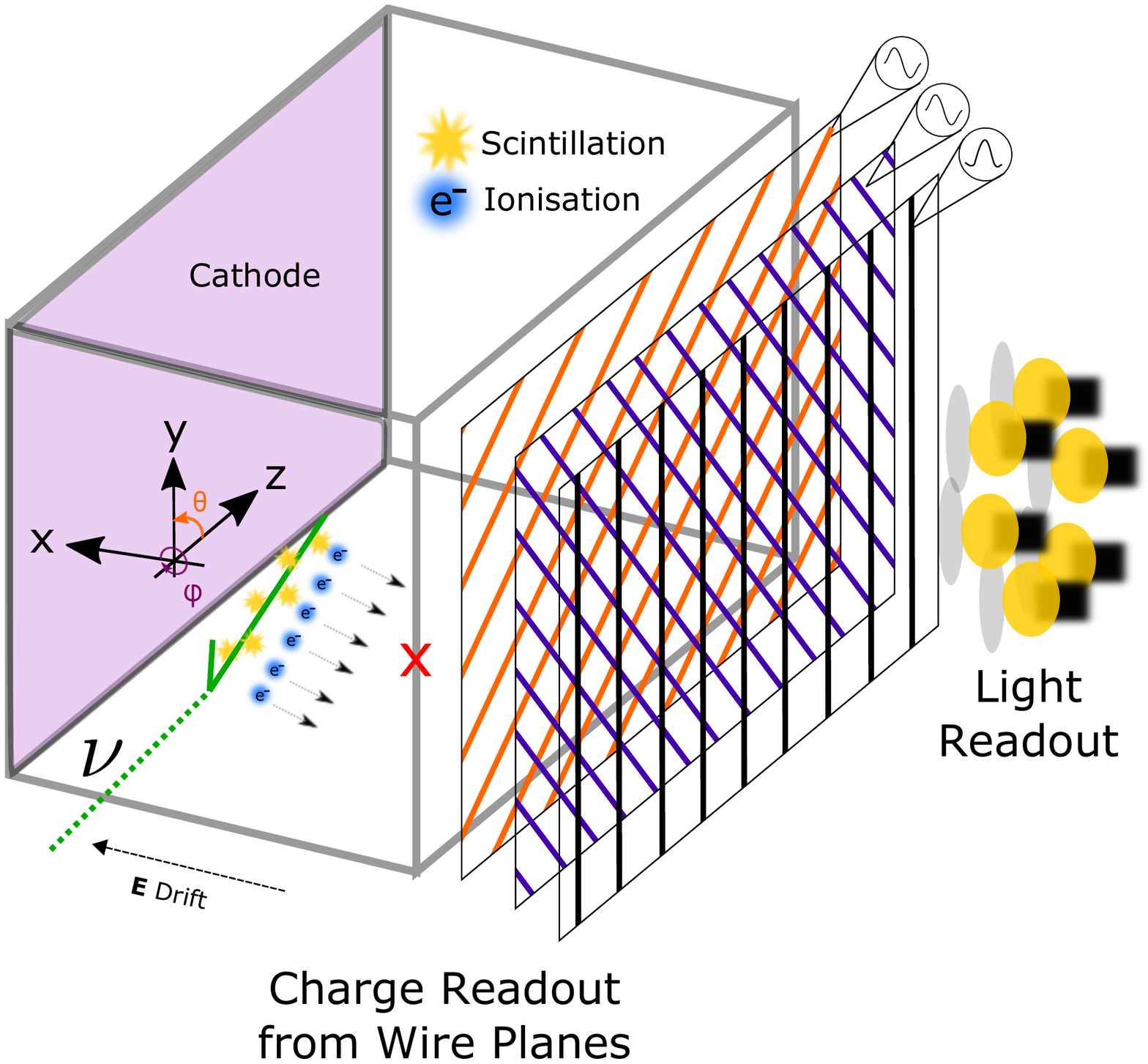}
    \caption{A diagram of a LArTPC with the coordinate system used in this analysis. The $z$ coordinate points in the direction along the Booster Neutrino Beam, MicroBooNE's primary neutrino beam (see text for details); $y$ in the upwards direction of the TPC; and $x$ from the three wire planes (anode) to the cathode (colored in purple). The red cross symbol marks the coordinate system origin. The angle $\theta$ is defined as the angle off the $z$-axis and the angle $\phi$ is defined as the angle in the $xy$ plane with $\phi=0^{\circ}$ pointing toward the cathode.}
    \label{fig:coordsystem}
\end{figure}

\section{The MicroBooNE Experiment}\label{sec:microboone_expt}

The MicroBooNE experiment is a LArTPC with an 85~tonne active mass, housed inside of a stainless-steel cryostat. The TPC has dimensions of 2.56~m (width, $x$), 2.30~m (height, $y$), and 10.37~m (length, $z$). Figure~\ref{fig:coordsystem} shows a diagram of the LArTPC together with the coordinate system used in this analysis. Here, we only focus on the elements of the detector crucial to this analysis. A more in-depth description of the MicroBooNE experiment is given in Ref.~\cite{microboone_experiment}.

\begin{figure}[tbh!]
    \center
    \includegraphics[width=\linewidth]{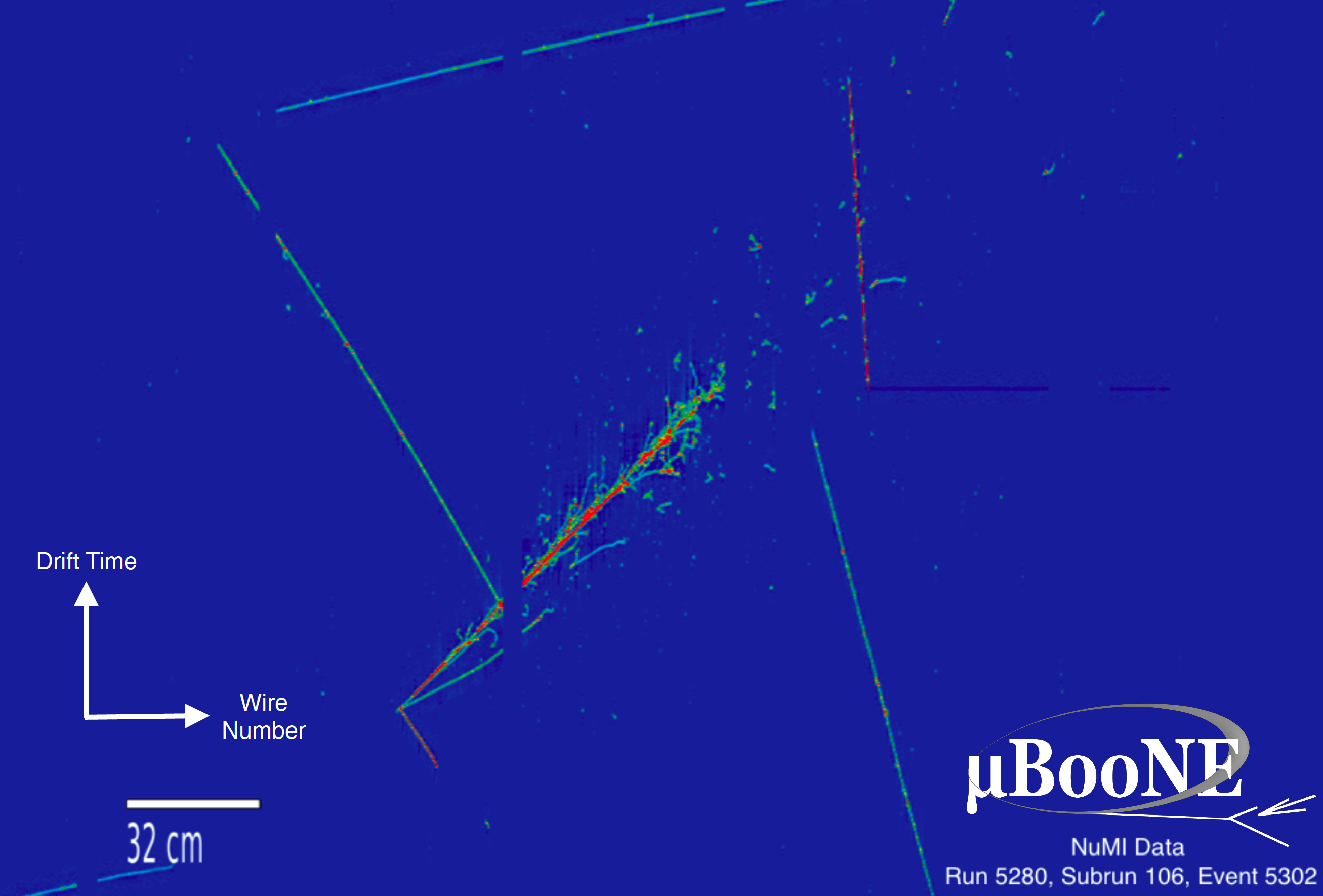}
    \caption{A display of a selected electron neutrino candidate recorded by the MicroBooNE detector using the NuMI beam alongside a number of cosmic ray tracks. The horizontal direction represents the wires on the collection plane and the vertical direction represents the electron drift time. Colors represent the amount of charge deposited on the wires. The gaps in some of the cosmic ray tracks and the electromagnetic shower are due to unresponsive wires.}
    \label{eventdisplay1}
\end{figure}

\begin{figure*}[th!]
    \center
    \includegraphics[width=\linewidth]{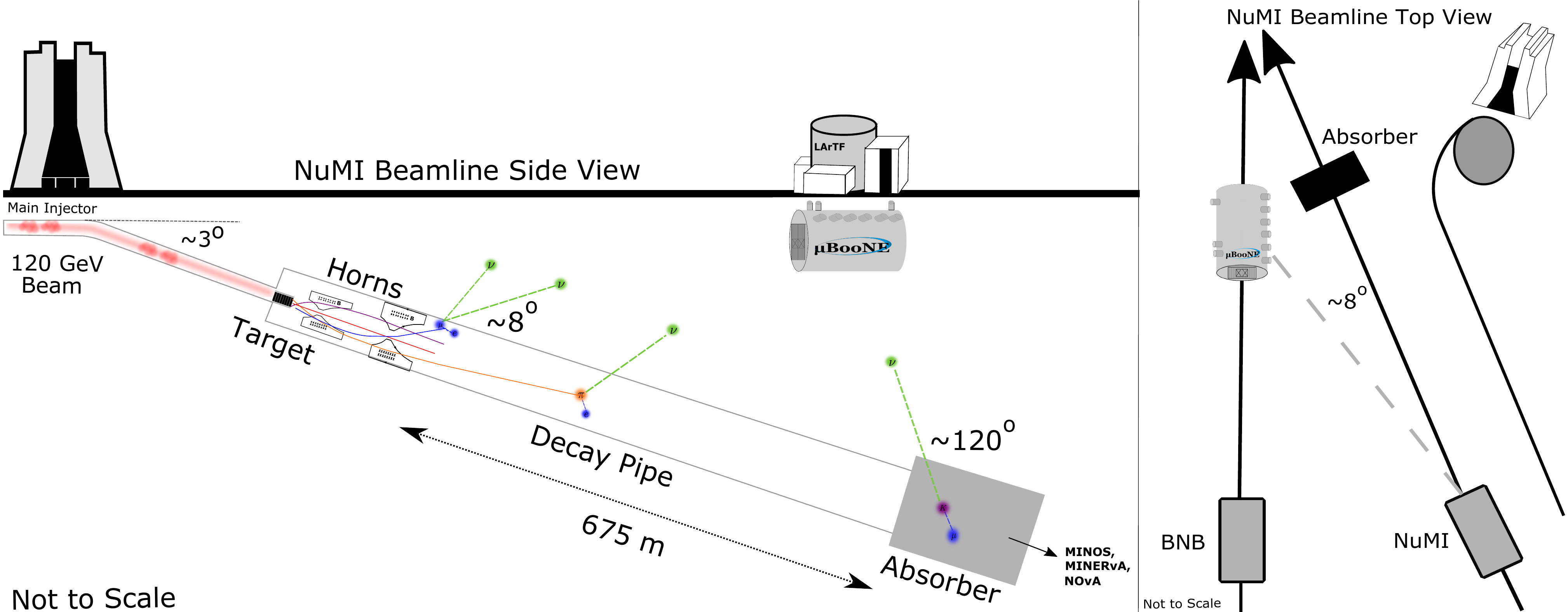}
    \caption{The position of the MicroBooNE detector relative to the NuMI neutrino beam target with views projected to the side and above. The NuMI beamline is angled 3$^{\circ}$ downwards and the distance of the NuMI target to MicroBooNE is approximately 679~m. The flux of neutrinos at MicroBooNE covers angles ranging from 8$^{\circ}$ to 120$^{\circ}$ relative to the NuMI beamline direction. }
    \label{fig:numidiagram}
\end{figure*}

Charged particles traversing the liquid argon ionize and excite the argon atoms and generate free electrons and scintillation light along their path. This scintillation light is detected by 32 photomultiplier tubes (PMTs) located behind the anode. A section of the full PMT system is depicted in Fig.~\ref{fig:coordsystem}. Each PMT gives a signal response within nanoseconds of the neutrino interaction which is significantly before the charge ionization signal is observed. To record a neutrino event, the MicroBooNE detector NuMI online trigger requires a scintillation light signal above 9.5~photo-electrons (PE) to be in-time with the accelerator beam spill window.

An electric field produced by a 70~kV drop over the 2.56~m drift distance attracts the free electrons towards an anode consisting of three planes of sensing wires. The free electrons induce a signal on the inner two wire planes (induction planes, with wires oriented at $\pm$60$^{\circ}$ from vertical) before being collected on the outer wire plane (collection plane, with wires oriented vertically). The signal on the wires provides position and calorimetric information for charged particles traversing the detector. Combining the information from the anode wire planes with timing obtained from scintillation light enables reconstruction of these interactions in 3D.

A candidate electron-neutrino interaction in the MicroBooNE detector recorded by the collection plane wires can be seen in Fig.~\ref{eventdisplay1}. The time needed for electrons to drift from the cathode to the anode is approximately 2.2~ms. In this time frame, multiple cosmic ray tracks cross the argon volume and can potentially contribute to the backgrounds of any neutrino analysis.
Cosmic ray tracks can be seen in Fig.~\ref{eventdisplay1} alongside the candidate electron-neutrino interaction.

MicroBooNE can detect neutrinos from the two neutrino beams produced at Fermilab. The detector is exposed to an on-axis flux from the BNB \cite{PhysRevD.79.072002}, and an off-axis flux of neutrinos from the NuMI beam \cite{numibeam}. The NuMI beam is created from collisions of protons accelerated to an energy of 120~GeV with a graphite target. These collisions start a particle cascade resulting in particles such as pions and kaons that can produce a neutrino from their decay. Particles of a particular electric charge from this cascade are focused by magnetic horns where the sign of the particles being focused depends on the direction of the electrical current applied to the horns. This analysis uses data from the NuMI beam in forward horn current mode which uses a horn current of +200~kA. This mode selects positively charged mesons and results in the on-axis flux being dominated by neutrinos. The majority of NuMI neutrinos interacting in MicroBooNE in the energy range used in this analysis originate at the beam target and arrive at the detector at an angle close to 8$^{\circ}$ relative to the NuMI beamline direction. The position of the MicroBooNE detector relative to the NuMI target is shown in Fig.~\ref{fig:numidiagram}. Due to the energy of the protons generating the beam and the position of MicroBooNE relative to the NuMI beamline, the NuMI neutrino flux at MicroBooNE has a composition of roughly 96\% $\nu_{\mu}+\bar{\nu}_{\mu}$ and 4\% $\nu_{e}+\bar{\nu}_e$ for energies above 250~MeV. The $\nu_{e}$ component is a factor of 10 larger than in the BNB making it an excellent source of electron neutrinos.

\begin{figure}[bth!]
    \center
    \includegraphics[width=0.45\textwidth]{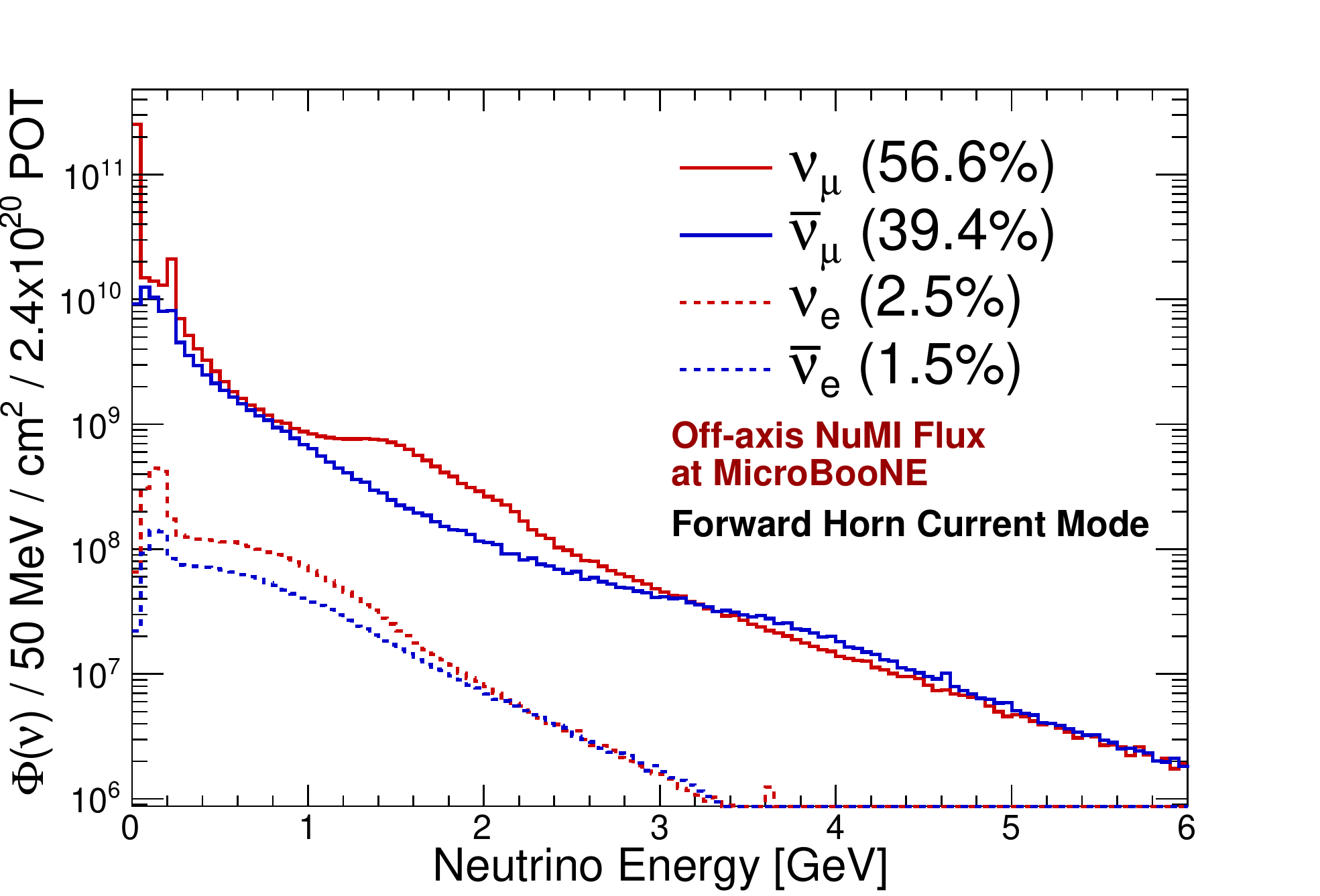}
    \caption{The NuMI beam neutrino flux incident on the MicroBooNE detector during NuMI forward horn current running. The percentages shown are calculated by applying a 250~MeV threshold on the neutrino energy.}
    \label{numiflux}
\end{figure}

The NuMI beam neutrino flux at MicroBooNE for each neutrino flavor is shown in Fig.~\ref{numiflux}. Each NuMI accelerator beam spill delivers $\sim10^{13}$ protons on target (POT) over a duration of 9.6 $\mu$s. Each accelerator spill consists of six proton batches. To increase the neutrino intensity the accelerator complex can be run in slip-stacking mode, doubling the proton intensity in some batches \cite{Brown:2013idd}. Between October 2015 and July 2016, in the course of its first year of data-taking, MicroBooNE collected $2.4\!\times\!10^{20}$~POT of NuMI beam data while the NuMI beam operated in forward horn current mode. The majority of these data was collected with the NuMI beam in a 4+6 slip-stacking configuration which means the first four out of the six proton batches have double the usual intensity. A smaller fraction of the NuMI data taken during this period also contain 5+6 and 6+6 slip-stacked data. The NuMI simulation used by MicroBooNE consistently assumes 6 batches (i.e. no slip-stacking). Due to the low neutrino interaction rate at MicroBooNE, multiple interactions in one spill are rare. We thus scale the simulated events to match the integrated data exposure, neglecting the sub-percent effects of pile-up.

\section{Simulation and Reconstruction} \label{sec:sim_and_reco}

The simulation and reconstruction of neutrino events from the NuMI beam in MicroBooNE is a complex set of steps that needs to account for both neutrinos and cosmic rays interacting within the detector. 
The proton interactions at the NuMI target, the meson re-interactions in the NuMI beamline, and the resulting flux of neutrinos are simulated using a custom simulation package, \texttt{FLUGG}, developed by the MINOS collaboration. This package combines \texttt{FLUKA} \cite{Fasso:1993kr} to model the particle interactions and \texttt{Geant4} \cite{Agostinelli:2002hh} to model the beamline geometry \cite{Campanella:683723}. The flux prediction additionally uses the \texttt{PPFX} software \cite{AliagaSoplin:2016shs} which is also used by the NO$\nu$A and MINER$\nu$A experiments. \texttt{PPFX} uses data from fixed-target experiments to constrain the hadron production in the NuMI beamline.
To constrain the NuMI flux at MicroBooNE we use the \texttt{PPFX} thin target (targets of few interaction lengths) constraints. More details can be found in Ref.~\cite{AliagaSoplin:2016shs}. The neutrino flux is provided as input to the \texttt{GENIE} \cite{genie_manual} neutrino event generator \footnote{GENIE v2.12.2 was used in this analysis}. \texttt{GENIE} simulates the neutrino-argon interactions inside the MicroBooNE cryostat volume. 
In parallel with the neutrino generation, a spectrum of cosmic ray particles is simulated using the \texttt{CORSIKA} \cite{corsika} software package \footnote{CORSIKA version v7.4003 with constant mass composition model is used}. The resulting Monte Carlo simulated (MC) events are processed using the \texttt{LArSoft} \cite{Snider:2017wjd} software framework. \texttt{LArSoft} is an event-based toolkit to perform simulation, analysis and reconstruction of LArTPC events. 
In the simulation chain the neutrino interaction products and cosmic rays are propagated through the detector using \texttt{Geant4}, taking into account field inhomogeneities caused by space charge accumulation \cite{Abratenko_2020}. This is then fed into a detector simulation resulting in realistic waveforms on the anode sense-wires and PMTs. The version of the MicroBooNE simulation software used in this analysis does not include the effects of drift charge producing signals on neighbouring wires, known as dynamically induced charge~\cite{signal_process_part1,signal_process_part2}. This can impact the selection efficiency and backgrounds; its systematic effect is discussed in Section~\ref{sec:sys_uncertainty}. 


The data acquired by the detector are reconstructed using the same reconstruction chain as the MC generated interactions. The sense-wire waveform signals are deconvolved in one dimension with the electronics response measured for each wire plane, resulting in waveforms with charge deposits having a uni-polar signature on all wire planes. 
These charge deposits are then reconstructed as ``hits" and fed into the \texttt{Pandora} generic pattern-matching reconstruction framework \cite{pandora_uboone}, which uses topological and calorimetric information to reconstruct and classify charged particles as three-dimensional objects. \texttt{Pandora} separates these objects into ``tracks" (muon, proton and pion candidates) and ``showers" (electron and photon candidates), which are assembled into particle hierarchies based mainly on proximity and shared vertices. \texttt{Pandora} also identifies the candidate neutrino interaction point as the neutrino ``vertex". Calorimetric information is associated with the reconstructed 3D tracks and showers in the form of dE/dx: the amount of energy deposited per unit length.

The scintillation light signals acquired by the PMTs are also reconstructed. Light arriving at each PMT is translated into photo-electrons and assembled into optical hits. Optical hits from different PMTs are combined into ``flashes", which represent the total amount of light recorded from a single neutrino interaction or a cosmic ray.
Flashes are characterized by position, given by the PE-weighted positions of the included PMTs, and time.
The scintillation light is 
used to determine the time of each interaction reconstructed in the TPC.



The $2.4\!\times\!10^{20}$~POT dataset used in this analysis 
 corresponds to 6,361,077 NuMI accelerator beam triggers with 734,221 of these passing the NuMI online trigger. We refer to these events as beam-on data.
We compare these to a sample of 728,500 
\texttt{GENIE} generated $\nu_\mu$, $\bar{\nu}_\mu$, $\nu_e$, $\bar{\nu}_e$ interactions inside the cryostat of MicroBooNE corresponding to $1.83\!\times\!10^{21}$~POT.
Each of these events also contains \texttt{CORSIKA} generated cosmic rays.
Neutrinos that interact within and outside the cryostat walls of the MicroBooNE detector can produce daughter particles which can travel inside the cryostat and produce enough light to pass the NuMI online trigger. These are known as out-of-cryostat interactions. We utilize a sample of 407,926 \texttt{GENIE}  $\nu_\mu$, $\bar{\nu}_\mu$, $\nu_e$, $\bar{\nu}_e$ interactions generated within and outside the MicroBooNE cryostat walls (with daughter particles that travel inside the cryostat) to estimate this beam-induced background. Each out-of-cryostat interaction is combined with \texttt{CORSIKA} generated cosmic rays in the MicroBooNE cryostat. The out-of-cryostat sample corresponds to $1.42\!\,\times\!\,10^{21}$~POT. All the MC samples generated are normalized to the total POT of the beam-on data sample. 

Not all accelerator spills result in a neutrino interaction in MicroBooNE. In many cases, the detector reads out exclusively cosmic rays in-time with the beam window. To characterize these readout triggers when no neutrino is present, a dedicated sample of 6,264,334 triggers was collected explicitly when the beam was off. This sample is normalized to the number of triggers for the beam-on data.

\section{Selection of Inclusive Charged-Current $\nu_e$-like Interactions}\label{sec:selection}
\input{cutSelectionfromIntNote}

\section{Flux-Averaged Inclusive $\nu_e + \bar{\nu}_e$ CC Total Cross Section}\label{sec:xsec_measurement}

We employ the standard formula to extract the cross section $\left<\sigma\right>$:

\begin{equation}
    \left<\sigma\right> = \frac{N-B}{\epsilon \times N_{\text{target}} \times \Phi_{\nu_e + \bar{\nu}_e}}
\end{equation}

\noindent where $N$ is the total number of selected neutrino candidates, $B$ the number of selected background events, $\epsilon$ the selection efficiency, $N_{\text{target}}$ the number of target nucleons, and $\Phi_{\nu_e + \bar{\nu}_e}$ the integrated $\nu_e + \bar{\nu}_e$ POT-scaled flux. The number of target nucleons in the fiducial volume defined in this analysis is $3.47 \times 10^{31}$. The number of selected signal neutrinos, ($N - B$), in data is calculated to be $80.9 \pm 17.5$ where $N=214$ and $B = 133.1$ (see Table~\ref{cutflowtab}). The mean of the $\nu_e$ and $\bar{\nu}_e$ fluxes is 905~MeV, which is calculated by integrating the flux from 250~MeV. 

\begin{figure}
\center
\includegraphics[width=0.15\textwidth]{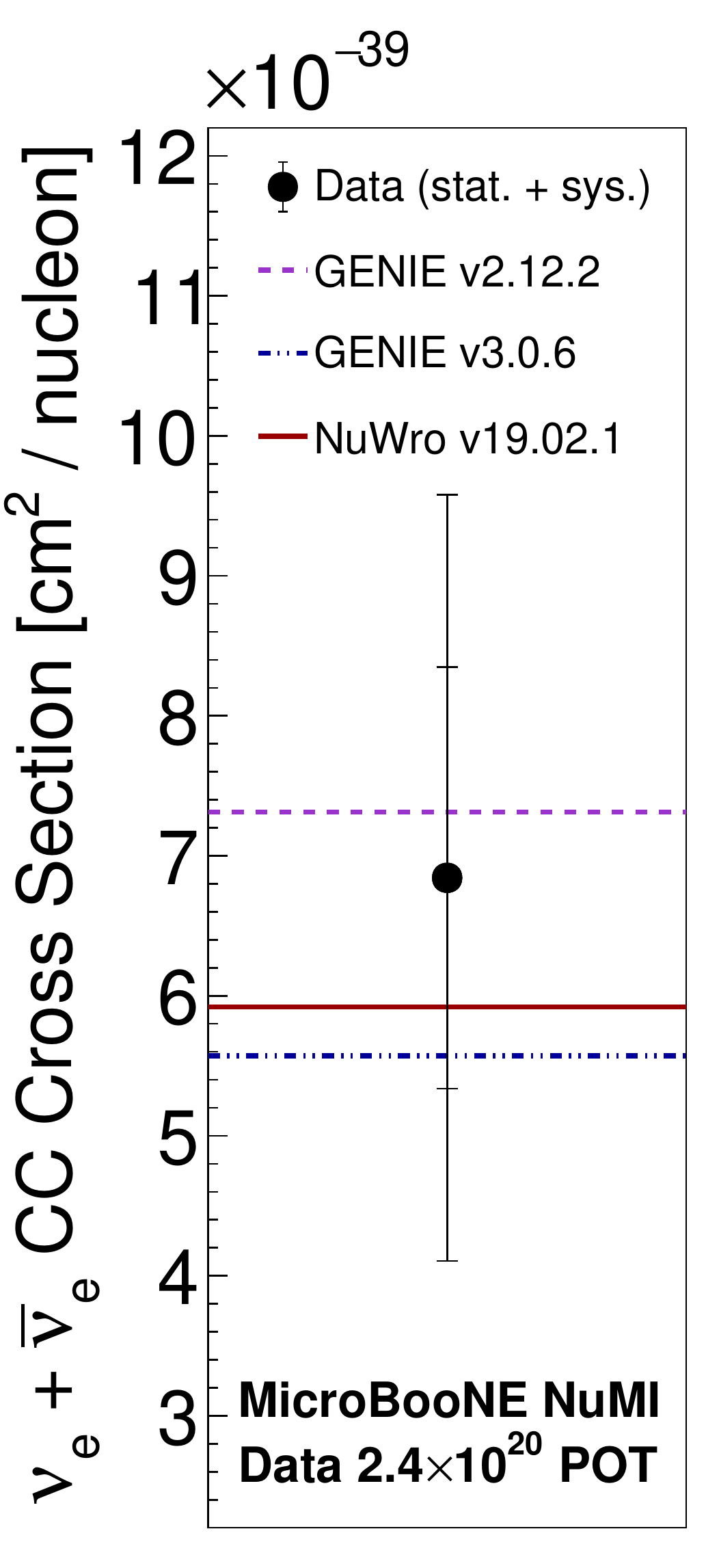}
\caption{ The extracted flux-averaged inclusive electron neutrino and antineutrino charged-current total cross section on argon compared to the predictions made by \texttt{GENIE} and \texttt{NuWro}. This measurement is for energies above 250~MeV and the average $\nu_e + \bar{\nu}_e$ flux at MicroBooNE above this threshold is 905~MeV. The measurement and predictions are in agreement within the statistical uncertainty.}
    \label{integrated_cross_section}
\end{figure}

The cross section, ${6.84\pm1.51\,\text{(stat.)}\times10^{-39}~ \textrm{cm}^2}$, is in agreement with the \texttt{GENIE} and \texttt{NuWro}~\cite{Golan:2012rfa} predicted values as seen in Fig.~\ref{integrated_cross_section}. The $N - B$ term is the leading contribution to the 22\% statistical uncertainty on the extracted cross section. We find similar agreement with \texttt{GENIE} v2.12.2 as ArgoNeuT does with \texttt{GENIE} v2.12.10c~\cite{Acciarri:2020lhp}. The theory predictions for the flux-averaged cross section between these versions of \texttt{GENIE} are equivalent.

\section{Systematic Uncertainties}\label{sec:sys_uncertainty}

The systematic uncertainties considered in the analysis arise from the simulation of neutrino interactions, propagation of secondary particles, detector response, and neutrino flux. 
The simulation of interactions on an argon nucleus is complex due to both the nature of the large nuclear target and the interplay of the different interaction modes in the 1~GeV energy region. 
\texttt{GENIE} is used to simulate the neutrino interactions with argon, using cross section models that depend on a number of parameters.
Estimates of the uncertainties on these parameters are provided by \texttt{GENIE}. These parameters in \texttt{GENIE} are simultaneously sampled 1000 times within their adopted uncertainties and are used to modify the simulated event rates.
This primarily modifies the background rates, though it does have a small impact on the signal efficiency.
For each of the 1000 variations, the measured cross section is re-calculated and the uncertainty is calculated as the standard deviation of the 1000 modified cross sections, leading to an uncertainty of 5\%.

The model parameter uncertainties provided by \texttt{GENIE} v2.12.2 are supplemented by considering alternative models within \texttt{GENIE} for charged-current quasi-elastic (CCQE) and meson exchange current (MEC) interactions - the dominant reaction mechanisms at MicroBooNE.
For low 4-momentum transfers (low-$Q^{2}$), the collective behavior of nucleons in the nucleus can lead to a suppression of the CCQE cross section. This physics effect is not included in our default \texttt{GENIE} model set. MEC interactions are simulated using an empirical model which does not include any associated uncertainties. An alternative CCQE model which includes suppression at low-$Q^{2}$ (calculated using the random phase approximation, or RPA) and a theory-driven MEC model for CC interactions \cite{PhysRevC.70.055503,Valencia1,Valencia2} are used to modify the default simulation.  Again, the cross section is recalculated with this modified simulation. This alternative model set introduces a 9\% change in the calculated cross section which is mostly driven by the effect of the alternative CCQE model on the efficiency (6\%).

Uncertainties for proton and charged pion re-interactions in argon are estimated by recalculating the survival probability as a function of momentum after modifying their cross sections within their uncertainty (conservatively estimated to be 30\%). The re-interaction cross sections for protons and charged pions are sampled simultaneously 250 times resulting in a new cross section in each case. Taking the standard deviation of these 250 modified cross sections results in an uncertainty of 2\%.

Systematic uncertainties originating from the detector modeling are evaluated using independent modifications to the detector simulation. Individual parameters in the underlying detector model are varied and the events re-simulated.
The measured cross section is recalculated using these modified simulations and the difference with respect to the central value is taken to be the uncertainty.
The uncertainties from the various detector parameters are added in quadrature, resulting in a 23\% uncertainty. 
The largest contribution to this uncertainty is the dynamically induced charge variation (16\%) where we use a simulation sample with this effect included. This variation affects the reconstruction of showers and greatly improves the data to MC agreement in variables such as the shower multiplicity and momentum. Future iterations of this analysis will include this effect in the default simulation.

\begin{table}[htb]
\center
\caption{Summary of systematic uncertainties on the cross section measurement in this analysis. The interaction uncertainty quoted here includes the \texttt{GENIE} (5\%), alternative CCQE and MEC models (9\%), and re-interaction (2\%) uncertainties. The beam flux uncertainty includes the hadron production (21\%) and beamline modeling (6\%) uncertainties. }
\label{systematics_table_summary}
\begin{tabular}{l  c}
\hline
\hline
Systematic Source & Relative Uncertainty [\%] \\
\hline
Interaction       & 10  \\
Detector Response & 23  \\
Beam Flux         & 22 \\
POT Counting      & 2   \\
Cosmic Simulation & 4   \\
Out-of-Cryostat Simulation   & 6  \\
\hline
Total & 34\\
\hline
\hline
\end{tabular}
\end{table}

The uncertainty in the flux prediction arises primarily from the modeling of the particle cascade following the proton-target collision. An alternate beamline simulation compatible with \texttt{PPFX} was run and reweighted in neutrino energy and angle to the NuMI beamline to match the nominal flux prediction from \texttt{FLUGG}. This reweighted flux was then modified by \texttt{PPFX} according to the hadron production uncertainties from data. Each time the flux was modified, the cross section was recalculated, including background subtraction and efficiency correction, giving a total uncertainty on the measurement from the hadron production of 21\%.

Additionally, we evaluate the uncertainties due to the modeling of the NuMI beamline by re-running the \texttt{PPFX}-compatible beamline simulation with parameters such as the target location, horn current, and beam spot size changed by their estimated uncertainties.
The combination of beamline uncertainties added in quadrature lead to a 6\% uncertainty on the measured cross section.
An additional uncertainty of 2\% is added to account for potential inaccuracies in the counting of POT from beamline monitors~\cite{2008PhDThesis68P,AliagaSoplin:2016shs}.


To estimate the uncertainty of the simulation of cosmic rays, we calculate the difference in selection rate of cosmic rays between a sample of simulated neutrino interactions overlaid on beam-off data and \texttt{CORSIKA} simulation. This difference is used to scale the selected \texttt{CORSIKA} cosmic rays in this analysis which varies the calculated cross section by 4\%. 

Simulation of out-of-cryostat interactions are dependent on many factors such as the geometry of the building around MicroBooNE, the density of various materials around the building and \texttt{GENIE} modeling of the neutrino interactions in these materials. Due to this large set of unknowns, the number of selected out-of-cryostat interactions is varied by 100\% and the cross section is recalculated. This gives an uncertainty on the cross section of 6\%.

Table \ref{systematics_table_summary} shows a summary of all the systematic uncertainties considered. We obtain a total uncertainty of 34\% with the flux and detector modeling being the most dominant. The final value of the electron neutrino and antineutrino CC total cross section on argon is therefore,

\begin{equation*}
    \left<\sigma\right> = 6.84 \pm 1.51\, \textrm{(stat.)} \pm 2.33\,\textrm{(sys.)} \times 10^{-39}~\frac{\textrm{cm}^{2}}{\textrm{nucleon}}.
\end{equation*}

This result is consistent with the \texttt{GENIE} prediction within statistical and systematic uncertainties as shown in Fig.~\ref{integrated_cross_section}.

\section{Conclusions}\label{sec:conc}

We presented the measurement of the flux-averaged inclusive electron neutrino and antineutrino charged-current total cross section on argon using the MicroBooNE detector and the NuMI beam at Fermilab. For $\nu_e + \bar{\nu}_e$ energies above 250~MeV and an average neutrino flux energy of 905~MeV calculated by applying this threshold, we find the cross section to be ${6.84\pm1.51\textrm{(stat.)}\pm2.33\textrm{(sys.)}\times 10^{-39}~\textrm{cm}^{2}/\textrm{nucleon}}$, which is in agreement with the predictions from \texttt{GENIE} and \texttt{NuWro}. This is the first such measurement performed in a large-scale LArTPC and the first one from a LArTPC placed on the surface. It is also the first measurement from an off-axis beam at MicroBooNE, with neutrinos arriving with a minimum angle of 8$^{\circ}$ relative to the NuMI neutrino beamline direction. Using the largest sample of electron-neutrino interactions on argon acquired to date, consisting of 214 selected $\nu_e$ and $\bar{\nu}_e$ CC events with a purity of 38.6\%, we demonstrate the electron-photon dE/dx separation power of LArTPCs using a fully-automated analysis chain. The measurement techniques presented here will be of immediate use for electron-neutrino appearance experiments such as the SBN program and DUNE.

\section{Acknowledgments}
This document was prepared by the MicroBooNE collaboration using the
resources of the Fermi National Accelerator Laboratory (Fermilab), a
U.S. Department of Energy, Office of Science, HEP User Facility.
Fermilab is managed by Fermi Research Alliance, LLC (FRA), acting
under Contract No. DE-AC02-07CH11359.  MicroBooNE is supported by the
following: the U.S. Department of Energy, Office of Science, Offices
of High Energy Physics and Nuclear Physics; the U.S. National Science
Foundation; the Swiss National Science Foundation; the Science and
Technology Facilities Council (STFC), part of the United Kingdom Research and Innovation;
 and The Royal Society (United Kingdom).  Additional support for the laser
calibration system and cosmic ray tagger was provided by the Albert
Einstein Center for Fundamental Physics, Bern, Switzerland.

 \bibliography{Bibliography}

\end{document}

%% file: cutSelectionfromIntNote.tex
\subsection{Event Classification}\label{event_classification}

\begin{table*}[htbp!]
\caption{A summary of the number of events in this analysis for data, simulated signal, beam background, cosmic MC background, and beam-off data background (scaled to the data POT/triggers).  The final two columns show the efficiency and purity at different stages of the selection.\label{cutflowtab}}
\center
\begin{tabular}{l  c  c  c  c  c  c  c }
\hline
\hline
Selection stage & Data & Signal & Beam Bgd. & Cosmic MC Bgd. & Beam-Off Bgd. & Efficiency [\%] & Purity [\%]\\
\hline
(1) Pre-selection  & 70691 & 632.1 & 7629.6 & 7736.4 & 52838.4 & 69.4 & 0.9 \\
(2) Flash matching & 11135 & 417.5 & 2160.8 & 613.7 & 6642.8 & 45.1 & 4.2 \\
(3) Vertex reconstruction quality & 7704 & 329.9 & 1462.4 & 457.3 & 4708.4 & 36.0 & 4.7 \\
(4) Shower hit threshold & 1889 & 276.9 & 509.5 & 82.6 & 725.0 & 29.9 & 17.4 \\
(5) Electron-like shower & 453 & 139.5 & 105.5 & 15.6 & 156.4 & 15.0 & 33.5 \\
(6) Final selection & 214 & 83.8 & 41.5 & 9.3 & 82.3 & 9.1 & 38.6 \\
\hline
\hline
\end{tabular}
\label{cutflowtab}
\end{table*}

We define our signal as a CC $\nu_e$ or $\bar{\nu}_e$ interaction inside a fiducial volume in the MicroBooNE detector above an (anti-)neutrino energy threshold of 250 MeV. This analysis is optimised towards high energies and therefore we set this threshold to exclude a region where the efficiency begins to rapidly decrease. Our signal events are identified by the presence of an electron or positron shower in the final state, regardless of the presence of additional particles. Because MicroBooNE is not able to differentiate electrons from positrons and, therefore, $\nu_e$ versus $\bar{\nu}_e$, the resulting selection contains both particles. As a consequence, we calculate the final cross section for a combination of $\nu_e$ and $\bar{\nu}_e$.

A pure selection containing $\nu_e$ and $\bar{\nu}_e$ CC interactions requires the use of several variables to remove any cosmic rays and other beam-induced backgrounds which are mis-reconstructed as showers. Due to the variety of interaction modes and detector effects, some interactions may be incorrectly classified, merged with other particles, partially reconstructed, or entirely unreconstructed. In order to study the signal efficiency and various background contributions, we classify events in the MC simulation as follows:

\boldsymbol{$\nu_e$} \textbf{CC}: $\nu_e$ or $\bar{\nu}_e$ interactions with an energy above 250~MeV with the primary interaction vertex inside the fiducial volume. This is our signal classification.

\boldsymbol{$\nu_e$} \textbf{CC Out-FV}: $\nu_e$ or $\bar{\nu}_e$ CC interactions whose primary interaction vertex is reconstructed inside the fiducial volume, while the true simulated vertex is located outside the fiducial volume but inside the MicroBooNE cryostat. As such, these are classified as background.

\textbf{Cosmic}: MC cosmic ray particles generated by \texttt{CORSIKA}~\cite{corsika} which are selected as the neutrino candidate. 

\boldsymbol{$\nu_{\mu}$} \textbf{CC}: MC generated particles originating from $\nu_{\mu}$ or $\bar{\nu}_{\mu}$ CC interactions. This background category includes all interaction topologies.

\textbf{NC}: MC generated particles from a neutrino neutral current (NC) interaction, including all topologies except those including $\pi^{0}$ in the final state.

\textbf{NC} \boldsymbol{$\pi^{0}$}: MC generated particles for a NC interaction with one or multiple $\pi^{0}$ in the final state. We classify these separately as the photons originating from $\pi^{0}$ decays can closely mimic electron showers.

\textbf{Out-of-Cryostat}: This category contains neutrino candidates originating from simulated neutrino interactions within and outside the cryostat walls.

\textbf{Beam-Off Data}: Any neutrino candidates originating from the sample of data collected when the beam was off fall under this background category. It contains exclusively cosmogenically produced activity.

 \subsection{NuMI $\nu_{\lowercase{e}}$~+~$\bar{\nu}_{\lowercase{e}}$ Selection}\label{numi_nue_selection}

We combine information from the NuMI beam extraction with the scintillation light recorded by the MicroBooNE PMTs and with TPC pattern recognition techniques. The selection does not target a specific part of the electromagnetic (EM) shower phase space, neither in angle nor energy. In order to reject beam and cosmic ray interactions that could mimic our signal, we apply a selection divided into six stages. These are listed in Table~\ref{cutflowtab}, together with the number of signal and background events surviving each stage.


The selection efficiency shown in Table~\ref{cutflowtab} is defined as the number of selected $\nu_e + \bar{\nu}_e$ CC interactions with an energy above 250~MeV in the fiducial volume divided by the number of simulated $\nu_e + \bar{\nu}_e$ CC interactions with the same energy threshold in the fiducial volume before any selection is applied. The selection purity is defined as the number of selected $\nu_e + \bar{\nu}_e$ CC interactions in the fiducial volume with an energy above 250~MeV divided by the total number of selected neutrino candidates (signal and background).

\subsubsection{Pre-selection}\label{section_fv_cut}

The goal of the first stage of the analysis is to identify events where a neutrino interaction happened inside a volume where they can be reliably reconstructed and has a flash coincident with the beam window. 

\texttt{Pandora} classifies each region of activity in the TPC as either a track or a shower. Our selection requires at least one reconstructed shower associated to the \texttt{Pandora} neutrino candidate. The \textit{leading shower} in the event is defined as the reconstructed shower object with the most charge deposition associated with it.

\begin{figure}[htb]
\includegraphics[width=\linewidth]{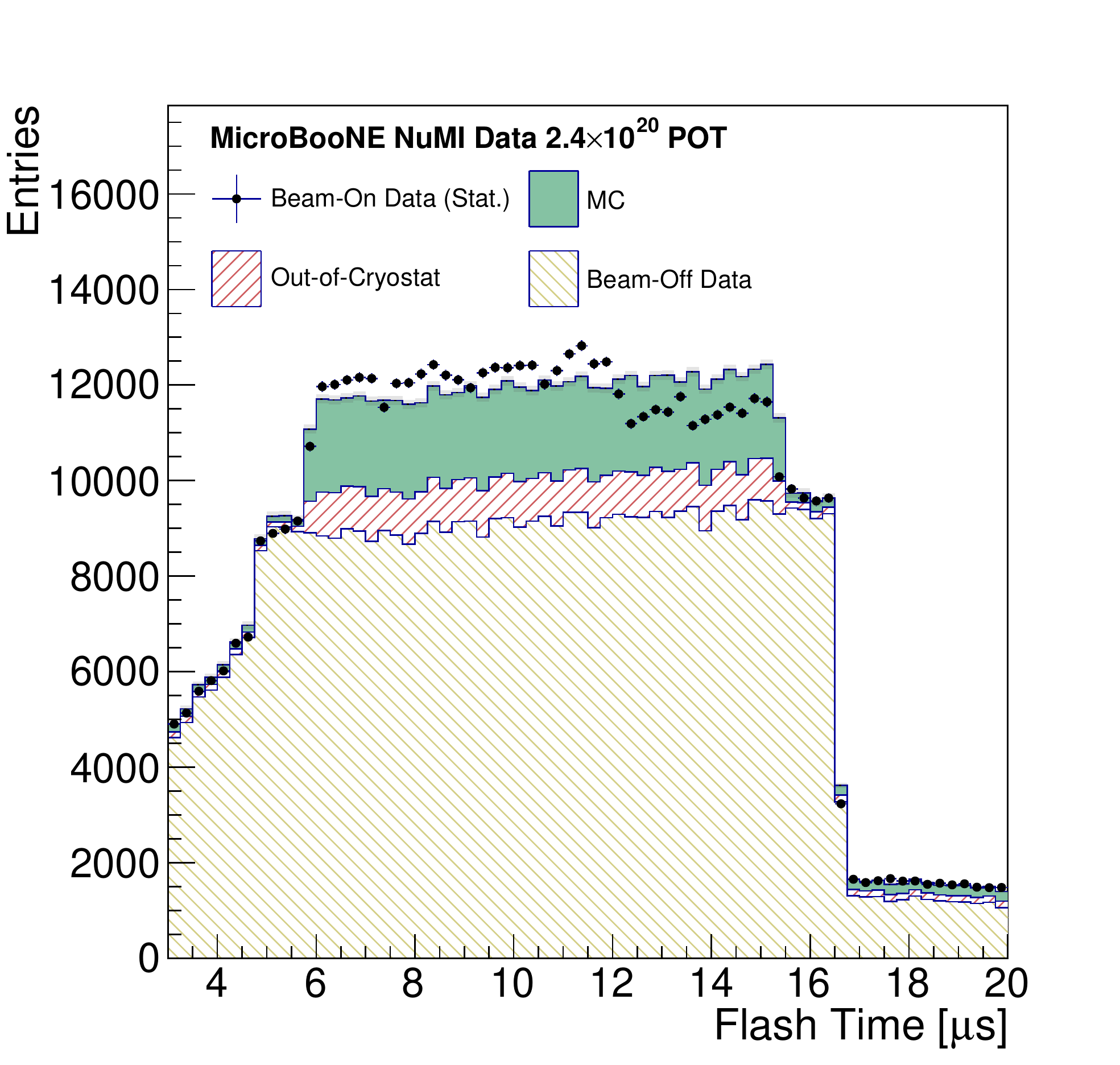}
\caption{Beam-on data (points) compared to prediction (MC + NuMI beam-off data) for the reconstructed flash time before the selection is applied. The initial flash time corresponds to the start of the MicroBooNE detector readout window. The NuMI beam spill occurs between 5.5 and 16.0~$\mu$s, where the greatest number of flashes are observed.}
\label{flash_time_data_everything}
\end{figure}

The distribution of the reconstructed flash time is shown in Fig.~\ref{flash_time_data_everything} for data versus the stacked prediction of MC + NuMI beam-off data. The NuMI beam spill window occurs between 5.5 to 16.0~$\mu$s. We reject flashes reconstructed outside of this window. 
The shoulder of the flash distribution on either side of the beam window is due to the NuMI online trigger gate being slightly wider than the NuMI beam spill window. This region is dominated by the beam-off data and is well-modeled. The shape between 3 and 4.5~$\mu$s is driven by flashes induced by cosmic activity that happen before the NuMI online trigger which have late scintillation light arriving inside the NuMI online trigger window. Flashes between 17 and 19~$\mu$s are generated mainly by argon late-light scintillation from interactions that happened during the beam window. The abrupt change in the number of flashes for beam-on data around 12 to 15~$\mu$s is a result of the 4+6 slip-stacking configuration of the NuMI beam. The MC is generated uniformly across the beam window such that the integral is equivalent to the integral of the beam-on data. This has no impact on normalization of the prediction to beam-on data because we normalize to the total POT delivered.

A large fraction of cosmogenic activity is caused by low energy neutrons and photons which deposit less energy on average than neutrinos. We therefore reject neutrino candidates where the total light signal observed by the PMTs is less than 50~PE. This is a highly efficient method of removing a significant number ($\approx5\%$) cosmic ray backgrounds of which most are low energy.


We define a fiducial volume in which we exclude 20~cm uniformly from all sides of the TPC, giving a total fiducial volume of $41.5$~m$^3$ (and a fiducial mass of 57.6~tonnes). Any neutrino interaction candidate with a reconstructed vertex outside of this volume is removed. This reduces the amount of selected out-of-cryostat and cosmic-ray interactions, and minimizes the impact of non-uniform electric field on the reconstructed tracks and showers near the cathode due to space charge accumulation~\cite{Abratenko_2020}.

The efficiency for the pre-selection is 69.4\%. This large initial loss in efficiency is primarily driven by the shower reconstruction performance, where the electron from the $\nu_e$ CC interaction is either mis-reconstructed or not reconstructed at all. 
At this stage, the purity is 0.9\%, as the selection is dominated by the cosmic ray background.

\subsubsection{Flash Matching}
A powerful method to reject cosmic ray backgrounds is to combine the TPC and light information. This is known as ``flash matching''.
In this analysis we calculate the distance from the position of the largest flash (i.e. the flash with the most PE) occurring inside of the beam window to the reconstructed \texttt{Pandora} vertices in 2D ($yz$ plane). We assume that the largest flash in the beam window was produced by the neutrino interaction. The distance, ${\Delta Y\!Z}$, is constructed as:

\begin{equation}
\Delta Y\!Z = \sqrt{(z_{\text{flash}} - z_{\text{tpc}})^{2} + (y_{\text{flash}} - y_{\text{tpc}})^{2}}~,
\end{equation}

\noindent where $z_{\text{flash}}$ and $y_{\text{flash}}$ are the reconstructed center of the largest flash and $z_{\text{tpc}}$ and $y_{\text{tpc}}$ are the reconstructed neutrino candidate vertex coordinates. 

Compared to neutrino interactions, cosmic-induced backgrounds (beam-off and \texttt{CORSIKA} MC) tend to have larger match distances when compared with the largest optical flash registered during the beam window. We select events using the 2D match distance taking into account the relative positions of the flash and the TPC interaction vertices in the $z$ coordinate. Since the neutrino interactions are usually forward going, the flash center should be downstream of the neutrino interaction. This motivates a tighter selection if the reconstructed TPC vertex is downstream of the flash position:

\begin{equation}
\begin{split}
z_{\text{tpc}} > z_{\text{flash}} \rightarrow \Delta Y\!Z < 60~\textrm{cm}~,\\
z_{\text{tpc}} < z_{\text{flash}} \rightarrow \Delta Y\!Z < 80~\textrm{cm}~.
\end{split}
\end{equation}

\subsubsection{Vertex Reconstruction Quality}

To mitigate backgrounds resulting from mis-reconstruction or incorrect particle hierarchy associations, we examine the distance of showers and tracks from the vertex, which is a metric of reconstruction  quality. This also removes background events where the leading shower originates from a photon. For example, in NC~$\pi^0$ events, the $\pi^0$  decays to two photons which pair convert after travelling some distance resulting in both showers being displaced from the vertex.

 


We remove any neutrino candidates where the leading shower is reconstructed with a start point further than 4~cm from the neutrino vertex. This requirement applies only to the leading shower to ensure an inclusive selection of $\nu_e$ CC topologies which produce showers distant from the neutrino interaction (e.g. $\nu_e$ CC $\pi^0$ production). If the neutrino candidate event includes reconstructed tracks, we additionally require that at least one track must start within 4~cm of the reconstructed neutrino vertex; this additional selection removes events with associated cosmic activity, while allowing for some mis-recontruction effects due to unresponsive wires. After applying these vertex quality variables, the purity increases to almost 5\%, see Table \ref{cutflowtab}.

\subsubsection{Shower Hit Threshold}

\begin{figure}[htbp]
\center
\includegraphics[width=\linewidth]{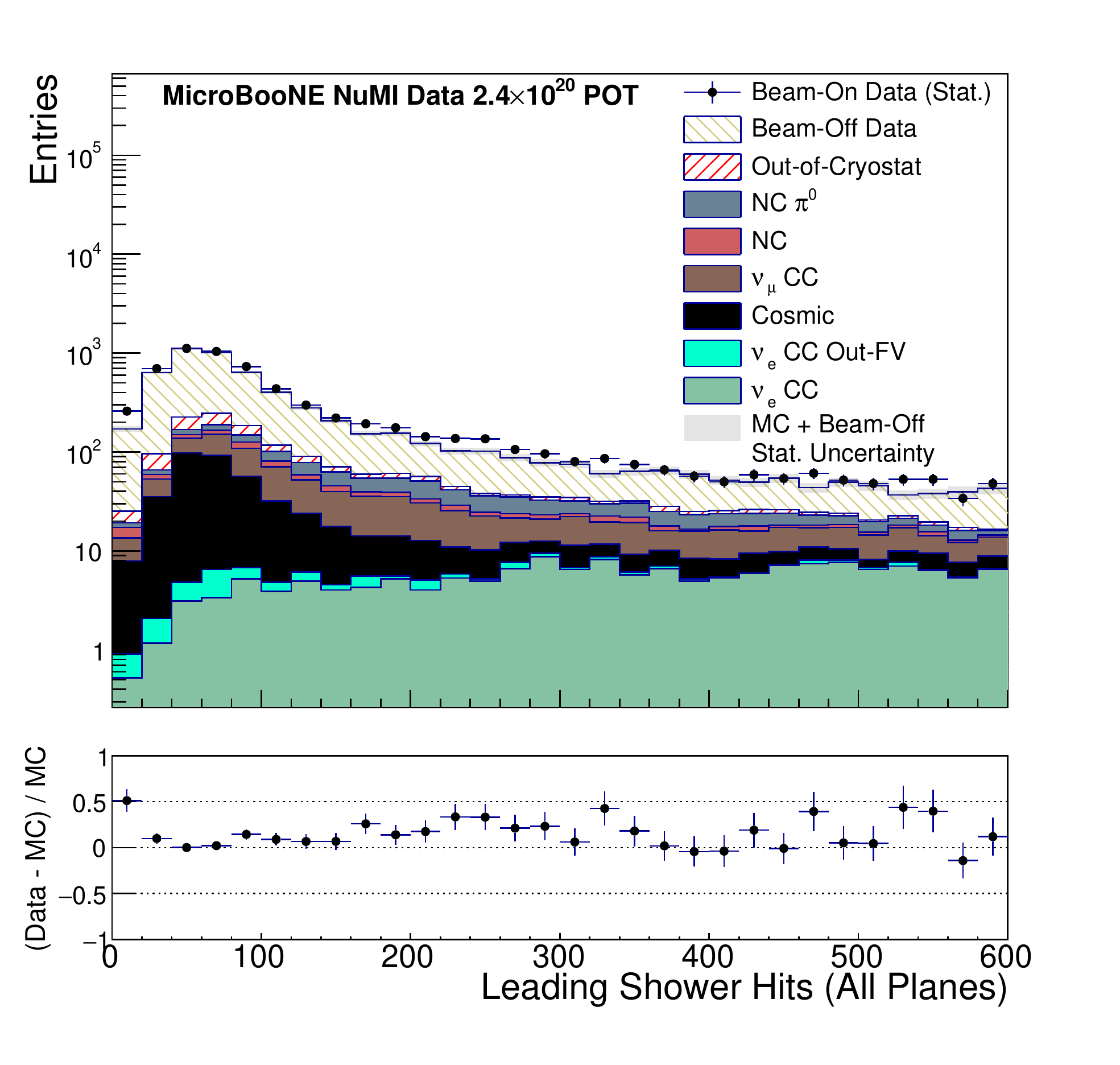}
\caption{Comparison of data (points) to the stacked prediction (MC + NuMI beam-off data) for the number of hits for the leading shower across all planes following the pre-selection, flash matching, and vertex reconstruction quality selection stages. }
\label{pre_hit_cut_total_hits_leading_shower_data}
\end{figure}

\begin{figure}[htbp]
\center
\includegraphics[width=\linewidth]{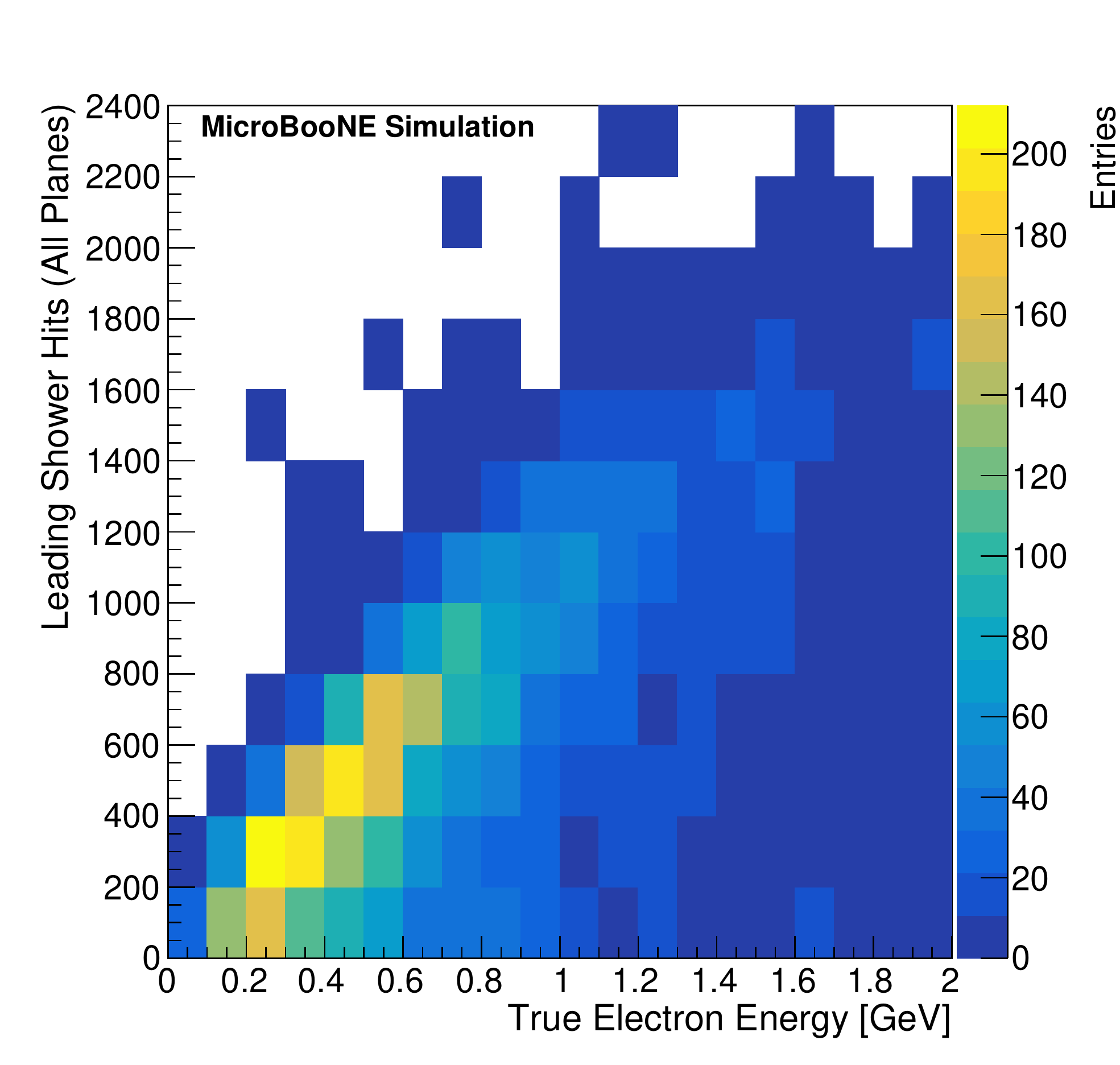}
\caption{Number of hits in the leading shower (all planes) against true electron and positron energy for all simulated $\nu_e$ and $\bar{\nu}_e$ interactions.
}\label{shwr_hits_ele_eng}
\end{figure}

The more hits that are associated to a shower, the easier it becomes to reconstruct its properties. Conversely, showers with very small numbers of hits are difficult to reconstruct precisely and are more likely to be affected by spurious charge depositions.
We integrate the number of hits for the leading shower across the three wire planes; the resulting distribution is shown in Fig.~\ref{pre_hit_cut_total_hits_leading_shower_data}. The majority of the backgrounds cluster at values below 200 total hits in the leading shower, mainly due to mis-reconstructed muon tracks or the reconstruction splitting larger showers into sets of smaller showers. Requiring greater than 200 hits for the leading shower is effective at removing these mis-reconstructed showers, however, it impacts low energy neutrinos as well as background events. Figure~\ref{shwr_hits_ele_eng} shows the relationship between the simulated electron energy and the leading shower hits. While the total number of hits correlates with the energy of the shower, this correlation is non-trivial. We observe that this selection requirement removes events in a bin of electron energy just above 0.2~GeV, but a large population of such events are maintained.



We also utilize the number of hits in the collection plane, which is typically the best-performing plane for reconstruction due to its high signal-to-noise ratio and is the plane used to calculate dE/dx at the start of the shower. Showers with few hits in this plane can lead to sizable uncertainties in the dE/dx measurement. Therefore, we require there to be at least 80 hits on the collection plane for the leading shower.


\subsubsection{Electron-like shower}

The presence of an electron shower is the identifying characteristic for determining whether an event was induced by an electron neutrino interaction. To isolate such events, we employ the key feature of LArTPCs: the ability to distinguish photon-induced showers from electron-induced showers using a combination of calorimetric and topological information (i.e. the measurement of dE/dx and the distance between the shower and  interaction vertex). The initial dE/dx for showers induced by an electron correspond to the minimum ionizing particle (MIP) value. Showers induced from photons will instead register higher values of initial dE/dx corresponding to double MIP ionization. This is due to the electron-positron pair from photon pair production, which is the dominant interaction mode for photons at the energies of interest. Before the pair production occurs, a photon does not ionize the argon. This can lead to an identifiable gap from the vertex which becomes another clear signature of background EM-showers.

Where the previous selection variables address mainly the backgrounds from cosmic rays and shower quality, requirements on the leading shower opening angle and the shower dE/dx further remove tracks that are mis-reconstructed as showers (mostly beam-off data and $\nu_{\mu}$ CC interactions) and limit the contamination from photon-induced showers originating from NC $\pi^{0}$ and $\nu_{\mu}$ CC $\pi^{0}$ interactions. 

\begin{figure}[htbp]
\center
\includegraphics[width=0.35\textwidth]{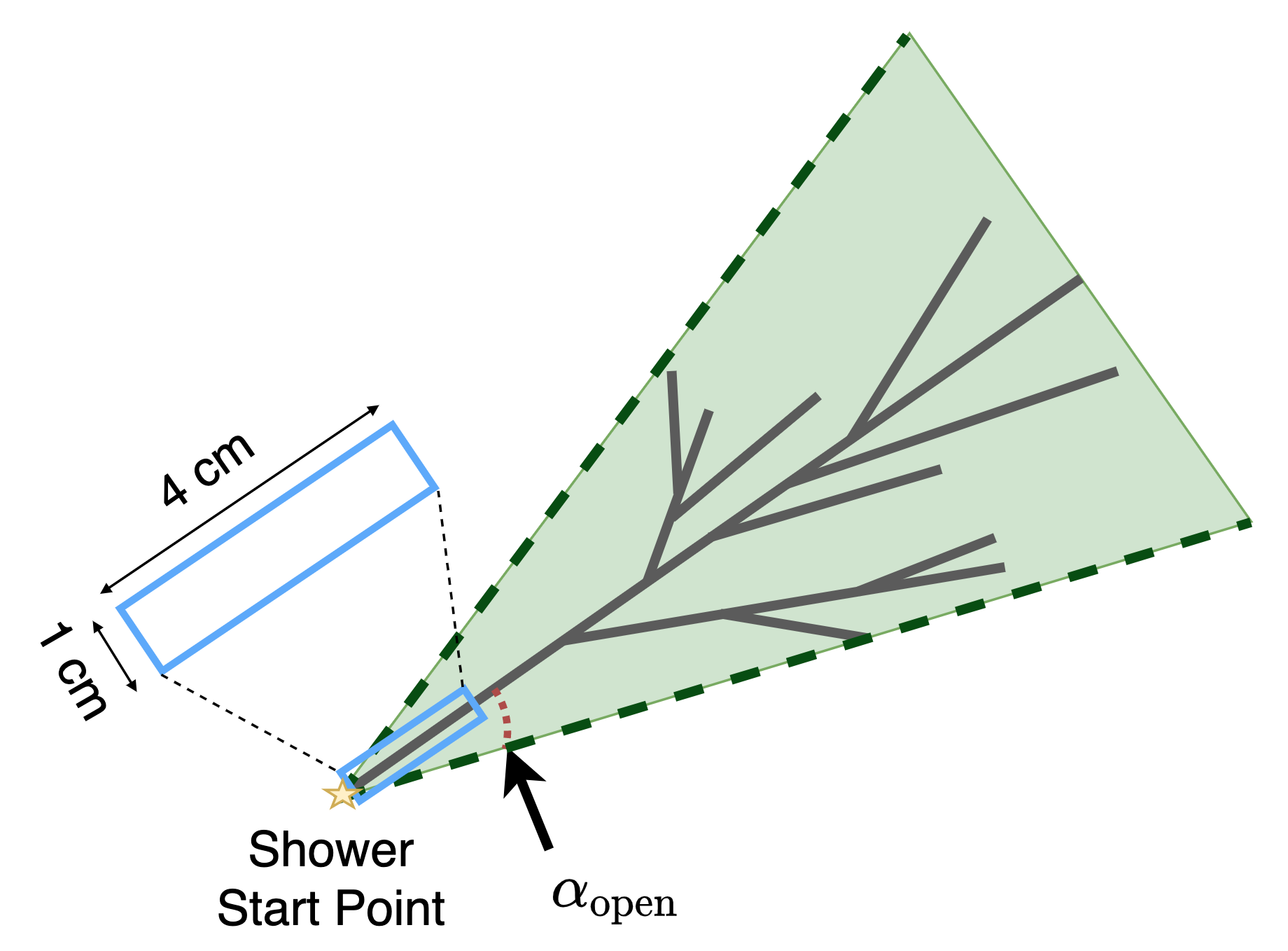}
\caption{Schematic showing the shower opening angle. The angle indicated by the dotted-red line shows the shower opening angle, $\alpha_{\text{open}}$. The median dE/dx calculation is performed using the charge deposition from the shower which falls within the box shown in blue.}
\label{shower_open_angle_schematic}
\end{figure}

\begin{figure}[htbp]
\center
\includegraphics[width=\linewidth]{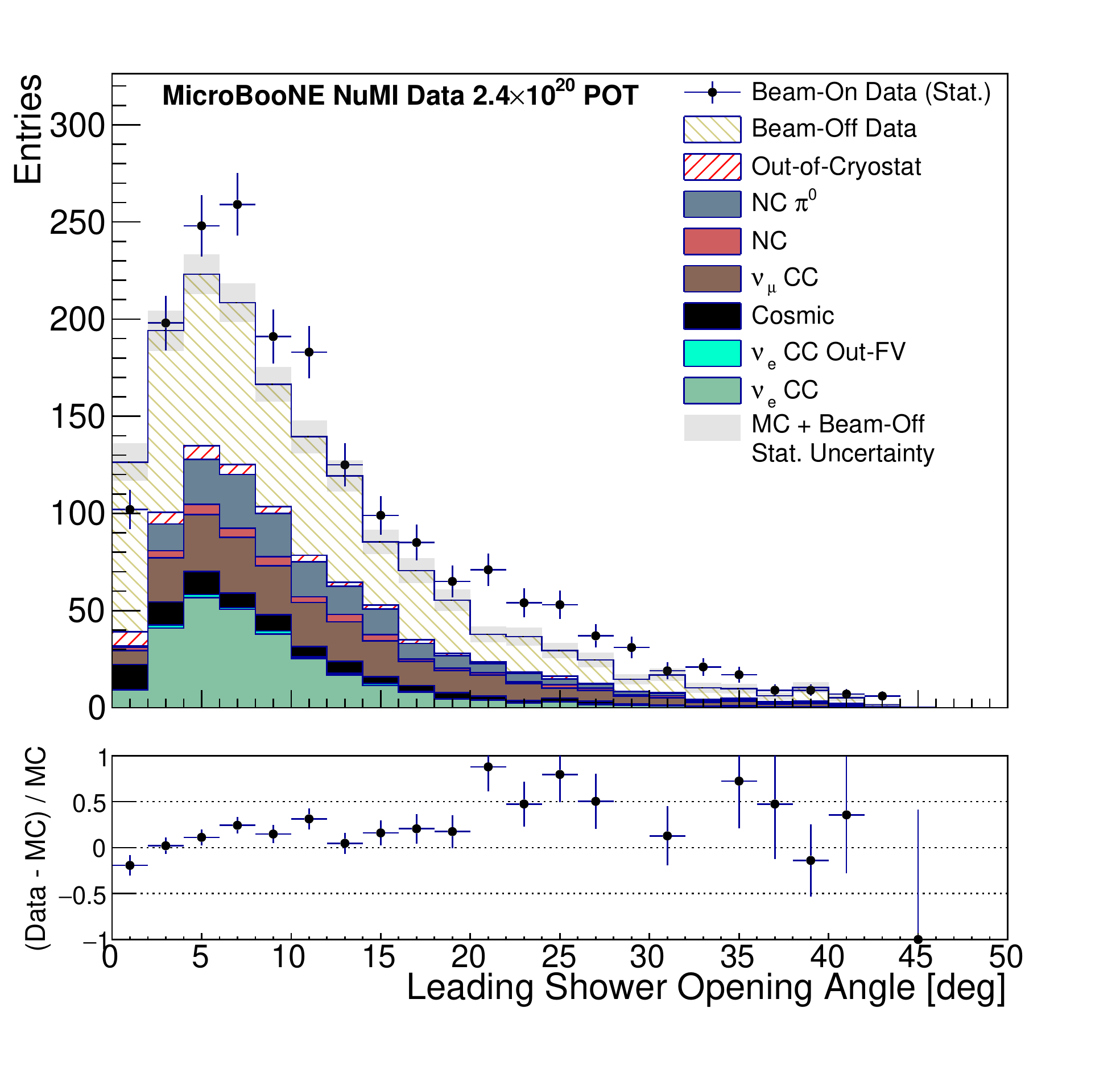}
\caption{Comparison of data (points) to the stacked prediction (MC + NuMI beam-off data) for the leading shower opening angle. Neutrino candidates pass the selection for leading showers between  3$^{\circ}$ and 15$^{\circ}$. }

\label{post_cuts_leading_shower_open_angle_data}
\end{figure}

The shower opening angle, $\alpha_{\text{open}}$, is calculated using a principal component analysis (PCA) of the 3D hit positions of the reconstructed shower and is given by the equation:

\begin{equation}
    \alpha_{\text{open}} = \tan^{-1}\left(\frac{\sqrt{\textrm{PCA}_{\textrm{secondary}}}}{\sqrt{\textrm{PCA}_{\textrm{principal}}}}\right),
\end{equation}

\noindent where $\textrm{PCA}_{\textrm{principal}}$ and $\textrm{PCA}_{\textrm{secondary}}$ are the lengths of the principal and secondary eigenvectors. Figure~\ref{shower_open_angle_schematic} gives an intuitive view of this angle on a schematic of a reconstructed shower. The opening angle is a powerful discriminator for cases where tracks have been mis-reconstructed as showers. For example, cosmic rays with broken tracks near the neutrino interaction can be mis-reconstructed as the leading shower resulting in a large opening angle. We select neutrino candidates where $\alpha_{\text{open}}<15^{\circ}$. In order to remove neutrino candidates with a topology which is more track-like than shower-like, we require a minimum opening angle of $\alpha_{\text{open}}>3^{\circ}$. Figure~\ref{post_cuts_leading_shower_open_angle_data} shows the data versus MC prediction for shower opening angle before this requirement is made.

One of the most powerful features of LArTPC technology is that its entire volume is an active calorimeter. This means that the energy loss of a particle can be calculated along its trajectory, enabling the use of dE/dx for particle identification. Using the dE/dx in the first few centimeters of a shower is a powerful tool to distinguish electron-induced from photon-induced showers, as was demonstrated by ArgoNeuT \cite{Acciarri:2016sli}. Selecting the median dE/dx value as a truer representation of the shower's deposition profile rather than the arithmetic mean mitigates effects from single outlier hits which could result from Landau fluctuations as well as mis-configured electronics, detector effects, or mis-reconstruction.

The calculation of the median dE/dx is performed by constructing a $1\!\times\!4$~cm$^2$ box starting at the reconstructed shower start point, shown in Fig.~\ref{shower_open_angle_schematic} and calculating the dE/dx for the collection plane single charge deposits along the start of the shower. The charge deposits are converted to an energy using the Modified Box model~\cite{Acciarri:2013met}.



\begin{figure}[htbp]
\center
\includegraphics[width=\linewidth]{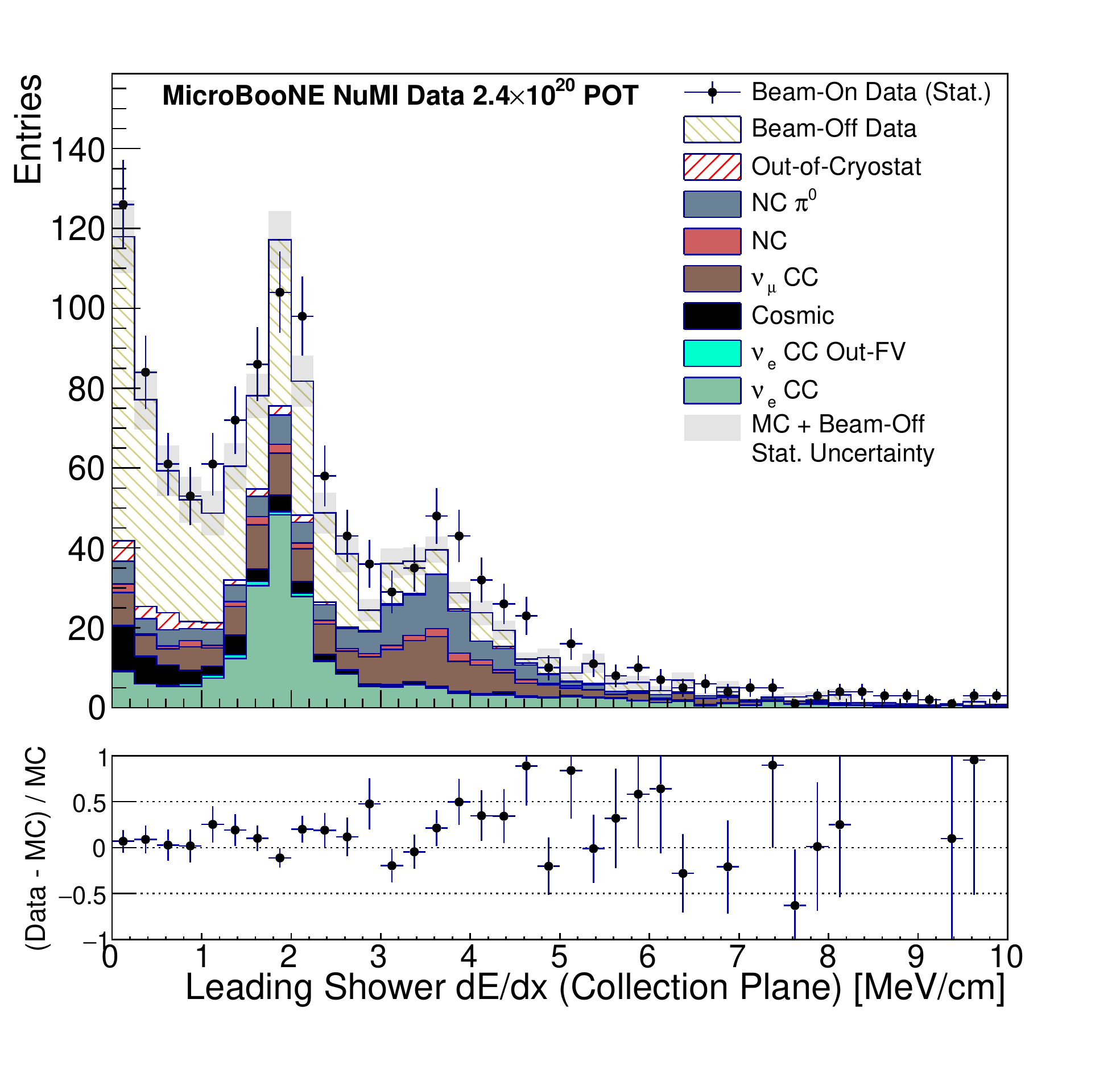}
\caption{Comparison of data (points) to the stacked prediction (MC + NuMI beam-off data) for the leading shower dE/dx (calculated as described in text) distribution after the shower opening angle selection requirement. Signal interactions are peaked at 2~MeV/cm and beam-induced backgrounds are mostly peaked around 4~MeV/cm. A large number of background events are found between 0 and 2~MeV/cm.}\label{post_cuts_dedx_cuts_data}
\end{figure}

Figure~\ref{post_cuts_dedx_cuts_data} shows the distribution of the calculated dE/dx on the collection plane wires for the leading shower using this method. The signal distribution peaks in the 2~MeV/cm region and large fractions of background lie to either side of the peak. Selecting those neutrino candidates whose dE/dx lies between 1.4~MeV/cm and 3~MeV/cm greatly increases the purity of the sample. As expected, the neutrino candidates containing photon-producing processes are centered around 4~MeV/cm. 

A notable feature in Fig.~\ref{post_cuts_dedx_cuts_data} is the large population of leading showers with a dE/dx of nearly 0~MeV/cm. This population is caused by tracks and showers that are nearly perpendicular to the beam direction ($60^{\circ}\!<\!\theta\!<\!120^{\circ}$) where it is challenging to measure dE/dx. In future analyses, this effect can be mitigated with the use of all three wire planes to measure dE/dx enabled by using methods such as 2D deconvolution as laid out in Refs.~\cite{signal_process_part1,signal_process_part2}.

\begin{figure}[t]
\center
\includegraphics[width=\linewidth]{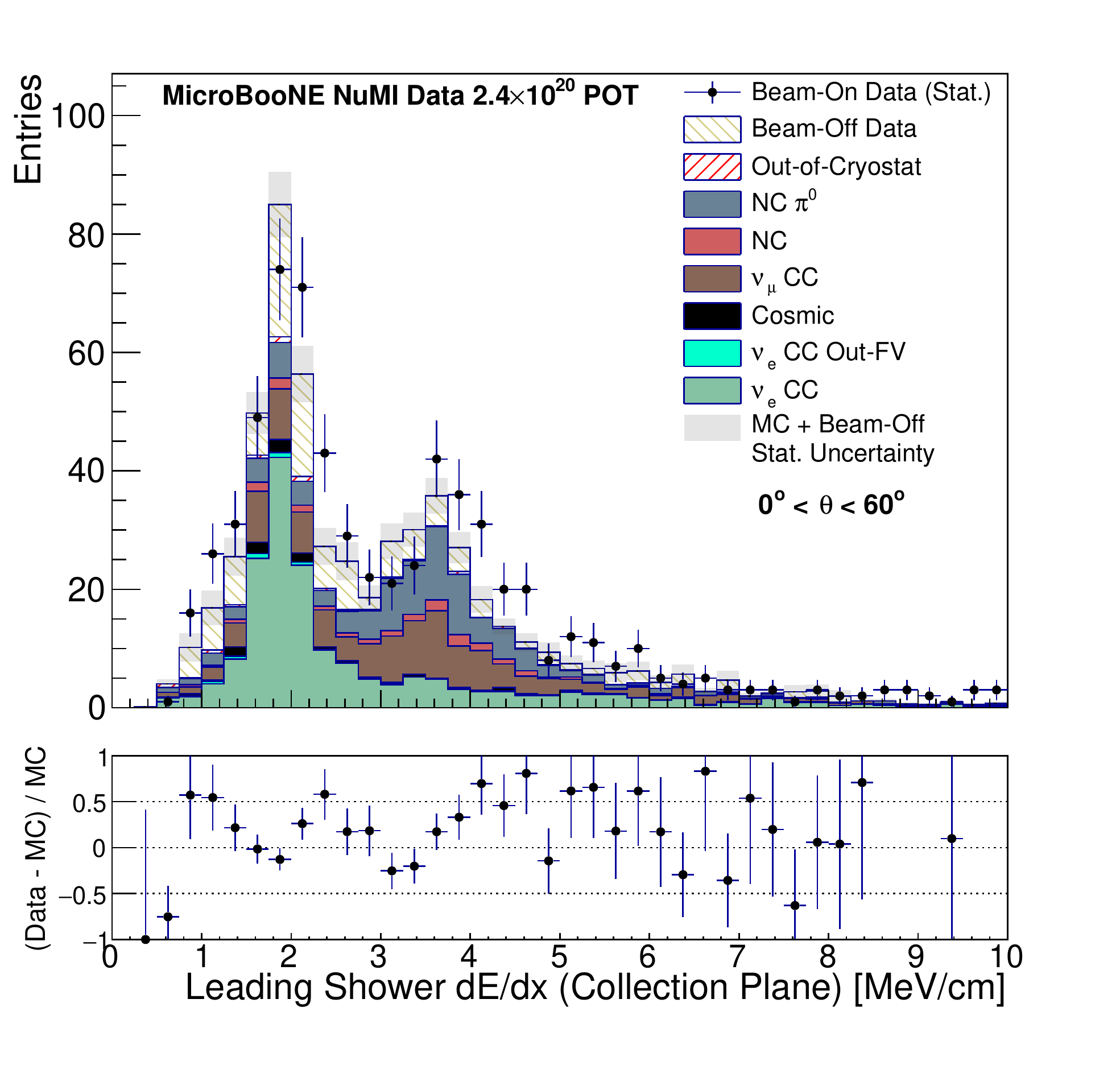}
\caption{Comparison of data (points) to the stacked prediction (MC + NuMI beam-off data) for the leading shower dE/dx for a slice of $\theta$ between 0$^{\circ}$ and 60$^{\circ}$ after the shower opening angle selection requirement. In this angular range, the leading showers all have a well reconstructed dE/dx and the signal peak is well-defined.} 
\label{post_cuts_dedx_theta_slice_1_data}
\end{figure}

Figure~\ref{post_cuts_dedx_theta_slice_1_data} shows the stacked data versus MC prediction where $\theta$ is between 0$^{\circ}$ and 60$^{\circ}$.  This slice of $\theta$ is the most populated region and has considerably higher purity than the rest of the phase space. As the dE/dx distribution at this angular slice includes showers running roughly perpendicular to the collection plane wires, a very small fraction of showers have an unphysically low dE/dx, which demonstrates the angular dependence of the dE/dx calculation in this analysis.

For values of $\theta$ between 60$^{\circ}$ and 120$^{\circ}$, where the dE/dx is not well reconstructed, both the neutrino interactions and the considerable cosmic ray background are peaked at dE/dx values closer to 0 MeV/cm. In the range of $\theta$ between 120$^{\circ}$ and 180$^{\circ}$, we expect relatively few electron-neutrinos and a high contamination of cosmic rays. However, in this sample, the majority of the leading showers have well reconstructed dE/dx close to 2 MeV/cm.

Given that the reconstructed leading shower direction can affect the calculation of dE/dx, using it may introduce an angular bias to the selection. 
However, it is an extremely powerful tool for removing cosmic and beam-induced backgrounds which dominate at low dE/dx, as well as photon backgrounds at higher dE/dx values. Future analyses can mitigate the angular dependence of dE/dx by using three plane calorimetry.

This selection stage is successful in removing a large fraction of photon-induced shower backgrounds. Given the importance of removing these backgrounds in electron-neutrino analyses we explore the performance of photon-rejection variables further in Section~\ref{sec:dEdx}.


\subsubsection{Final Selection}


Mis-reconstruction of the neutrino candidate can lead to associating physically uncorrelated showers with the interaction. To remove these cases, we require the distance between the sub-leading showers and the neutrino candidate vertex to be less than 22 cm. The scope of this selection requirement is to mitigate mis-reconstructed showers while retaining a sizable fraction of $\nu_e$ CC $\pi^{0}$ interactions ( 44\%  relative to the previous selection stage).


\begin{figure}[htbp]
\center
\includegraphics[width=\linewidth]{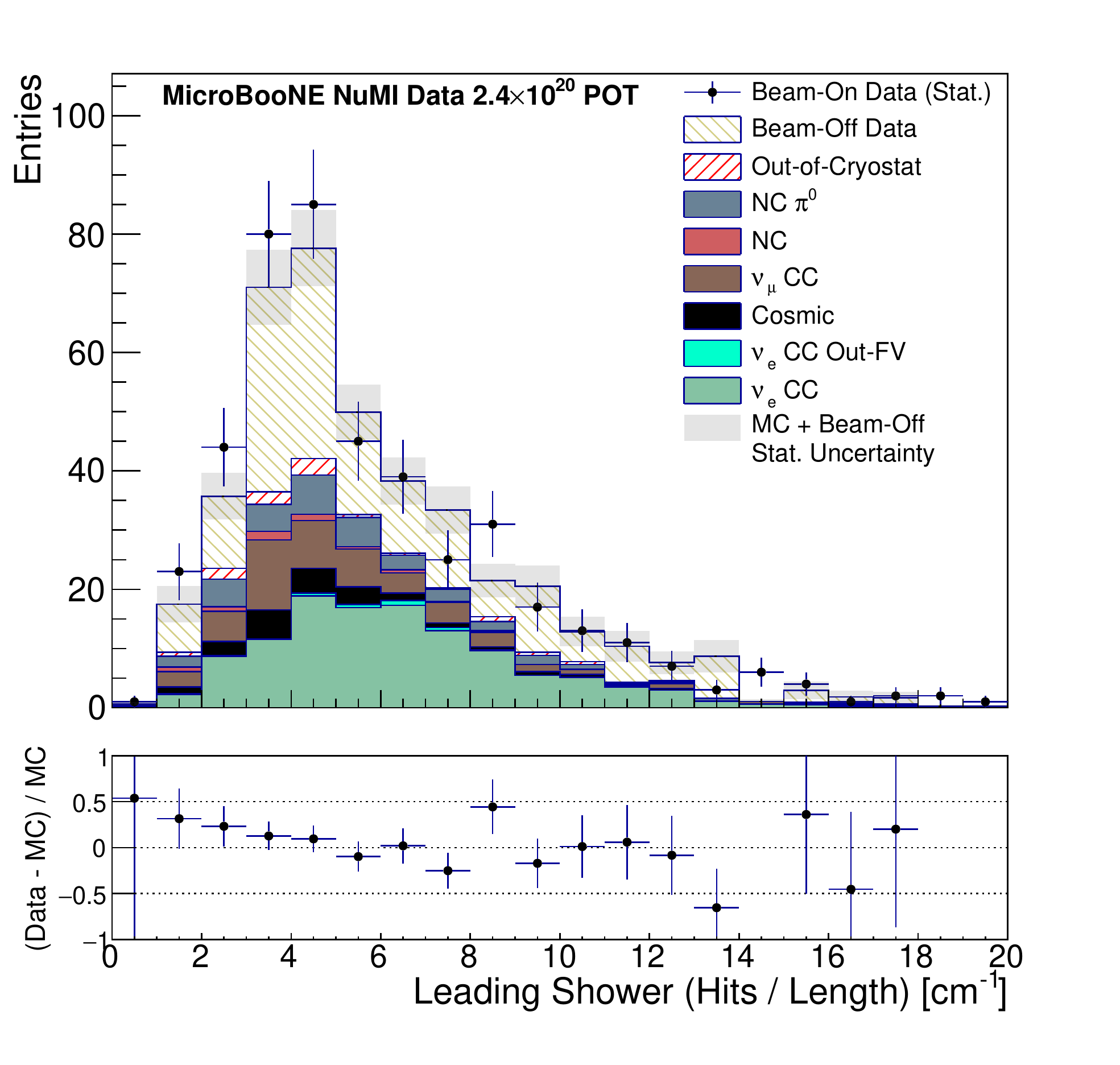}
\caption{Comparison of data (points) to the stacked prediction (MC + NuMI beam-off data) for the leading shower hit density variable after the selection requirement on the sub-leading shower distance to the vertex. The threshold is placed at 3~hits~per~cm where all neutrino candidates that have a leading shower hit density less than this value are removed.}\label{post_cuts_hitdensity_cuts_data}
\end{figure}

Further differentiating between shower and track objects is necessary in order to remove the cosmic ray interactions which dominate the currently selected sample. The object \textit{hit density} is calculated by summing the number of hits associated to the leading shower and dividing by its length. The hit density for the leading shower is shown in Fig.~\ref{post_cuts_hitdensity_cuts_data}, the cosmic rays largely populate the lower values of hit density. The effectiveness of this selection is particularly sensitive to the reconstruction of the transverse component of the shower; poorly reconstructed shower objects are removed by a selection on this variable. Placing a higher threshold on the hit density improves the selection purity, however the selection efficiency especially in the low energy signal region is impacted. A conservative threshold is placed at a hit density of 3 hits per cm.

\begin{figure}[htbp]
\includegraphics[width=\linewidth]{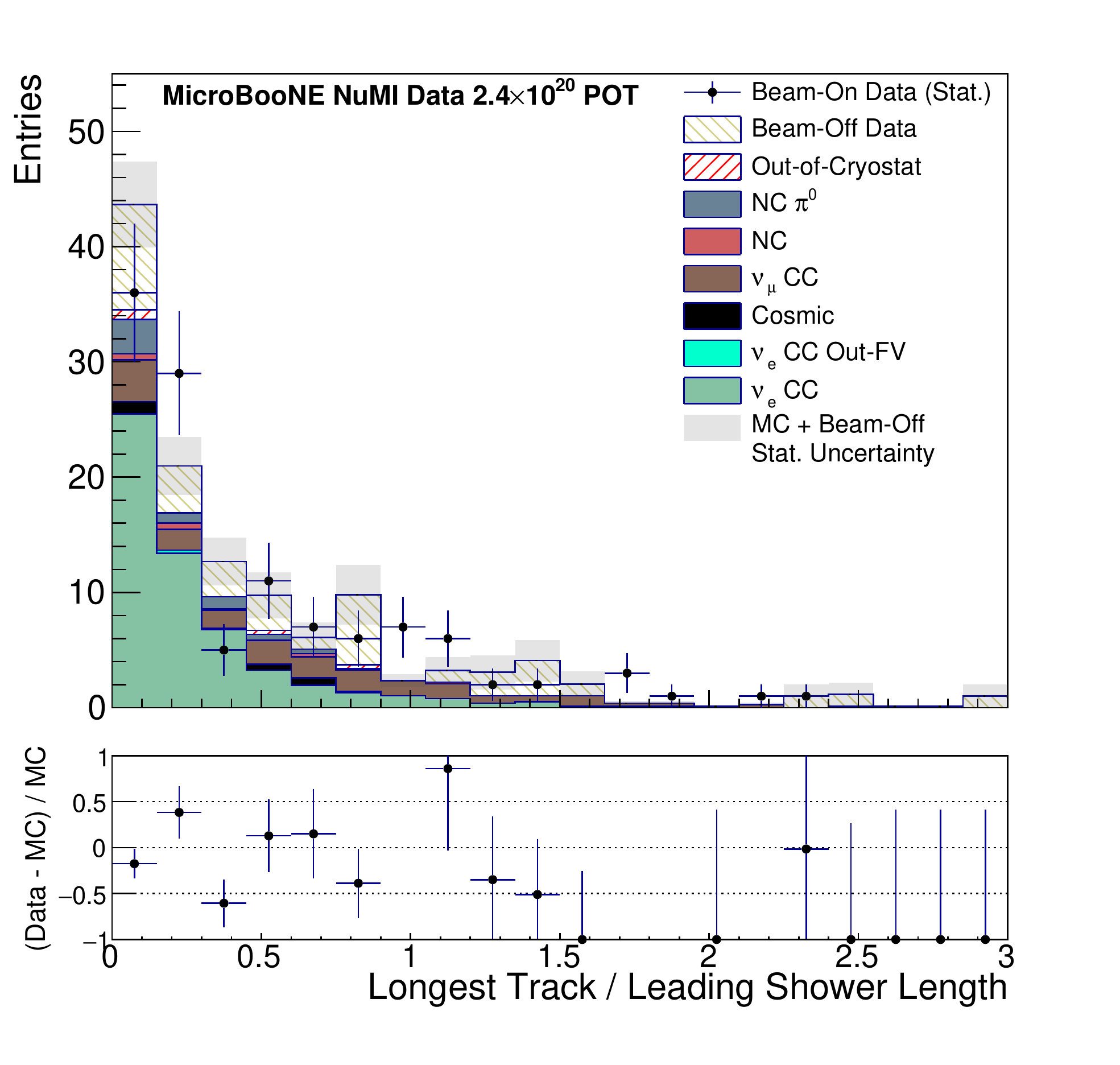}
\caption{Comparison of data (points) to the stacked prediction (MC + NuMI beam-off data) for the ratio between the longest track and the leading shower length following the selection requirement on the shower hit density. The threshold is placed at 1.0, where all neutrino candidates that have a track longer than the leading shower are removed.}\label{post_leading_shower_trk_lengths_data}
\end{figure}

The relationship between the length of the longest track and leading shower can be used to discriminate between $\nu_{\mu}$ and $\nu_e$ interactions.  For instance, a $\nu_{\mu}$ CC interaction typically contains a rather long muon track and any showers associated to the interaction are typically much shorter in length. Contrast this to a $\nu_e$ CC interaction, where the tracks produced are often of comparable length, or shorter than the leading shower length even in the presence of charged pions. Selecting on such a variable also removes cases where Michel electrons are produced by muon decays. The parameter shown in Fig.~\ref{post_leading_shower_trk_lengths_data} is defined as the length of the longest track in the neutrino candidate event divided by the leading shower length.  We select neutrino candidates whose ratio is below 1.0. 

To increase the purity of \textit{shower + track} topologies, we require that both the reconstructed start and end of the track are contained inside the fiducial volume. This requirement is particularly effective at rejecting long cosmic rays or muons which cross the fiducial volume and are associated to a shower inside the fiducial volume. This has a small impact on the signal selection and results roughly in a 3\% increase in purity. 


The final number of neutrino candidates remaining after each selection stage is shown in Table \ref{cutflowtab}. 


\subsection{Electron-Photon Separation}\label{sec:dEdx}

A key requirement of any analysis searching for electron neutrinos is the ability to differentiate electrons originating from $\nu_e$ CC interactions from photons originating from any backgrounds. The two main features that separate interactions containing electrons from those with photons are the dE/dx at the start of the shower and the distance between the shower and the interaction vertex. The latter is only well-defined when another charged particle is present at the interaction vertex. Electron-photon separation in a LArTPC has previously been demonstrated using a semi-automated reconstruction chain \cite{Acciarri:2016sli} and only leveraging the dE/dx. 
In this measurement, we demonstrate  for the first time both of the electron-photon separation techniques that the LArTPC technology offers using a fully automated analysis chain.

To examine the performance of the electron-photon separation variables, we isolate the dE/dx and shower vertex distance selection steps on the leading shower by moving them to the end of the analysis chain; this ensures that the upstream part of the selection chain identifies neutrino interactions with a well-defined leading shower. For this study, we additionally require the leading shower $\theta$ to be between 0$^{\circ}$ and 60$^{\circ}$. This focuses on the topologies unaffected by the absence of dynamically induced charge in our simulation chain and with dE/dx best reconstructed on the collection plane wires. The very good agreement between the  data and MC samples allows us to utilize the MC sample, which provides true information about the nature of the leading shower, to determine the power of the two separation methods.

After applying the $\nu_e+\bar{\nu}_e$ CC selection without the dE/dx and shower vertex distance selection steps, we obtain a sample of 1995 simulated neutrino events. In this sample, the true particle responsible for the leading shower is an electron in 48\% of cases, a photon in 39\% of cases with 13\% remaining for other particles. We then examine the individual and combined effect of applying the dE/dx and the shower to vertex distance selection requirements on these three groups. The value of dE/dx is required to be between 1.4 and 3~MeV/cm and the distance between the shower and the vertex to be less than 4~cm apart. The combination of these two requirements selects 59\% of electron neutrino events and rejects 81\% of photon backgrounds and over 61\% of other backgrounds. When applying the requirements individually, the dE/dx is the significantly more powerful method of rejecting events with photons removing 73\% of those backgrounds by itself compared to 28\% for the shower distance to vertex. It is also responsible for the bigger drop in our efficiency to select electrons: 35\% compared to 11\%. We also investigate the effect of the shower to vertex distance selection requirement on a subset of events with at least one candidate track present. For this sample, the selection requirement has an improved performance in rejecting photon backgrounds with 47\% rejected compared to 28\% for events where we do not require the presence of a reconstructed track. The summary of the performance for each selection requirement applied individually and combined can be found in Table~\ref{tab:ele-phot}.

\begin{table}[htb]
\center
\caption{Survival rate of a sample of 1995 neutrino events where the leading shower is classified as originating from an electron, photon, or other based on MC information. The EM shower selection row refers to the $\nu_e+\bar{\nu}_e$ CC selection without the dE/dx and shower to vertex distance selection requirements. The subsequent rows show the effect of the dE/dx and shower distance to vertex selection requirements applied individually and combined for this sample of events. The final row shows the performance of the shower vertex distance selection requirement applied to events with at least one candidate track present. \label{tab:ele-phot}}
\begin{tabular}{c  c  c  c  } 
\hline
\hline
Selection stage & Electrons & Photons & Other \\
\hline
EM Shower Selection & 951 & 771 & 273  \\
dE/dx (only) & 65\%  &  27\% & 52\%  \\
Shower-Vertex Dist. (only) & 89\% & 72\% & 73\% \\
Combined & 59\% & 19\% & 39\% \\
\hline
Shower-Vertex Dist.  & 89\% & 53\% & 64\% \\
(only, $\geq$ 1 track) & & & \\
\hline
\hline
\end{tabular}
\end{table}

\begin{figure}[tbh!]
\center
\includegraphics[width=\linewidth]{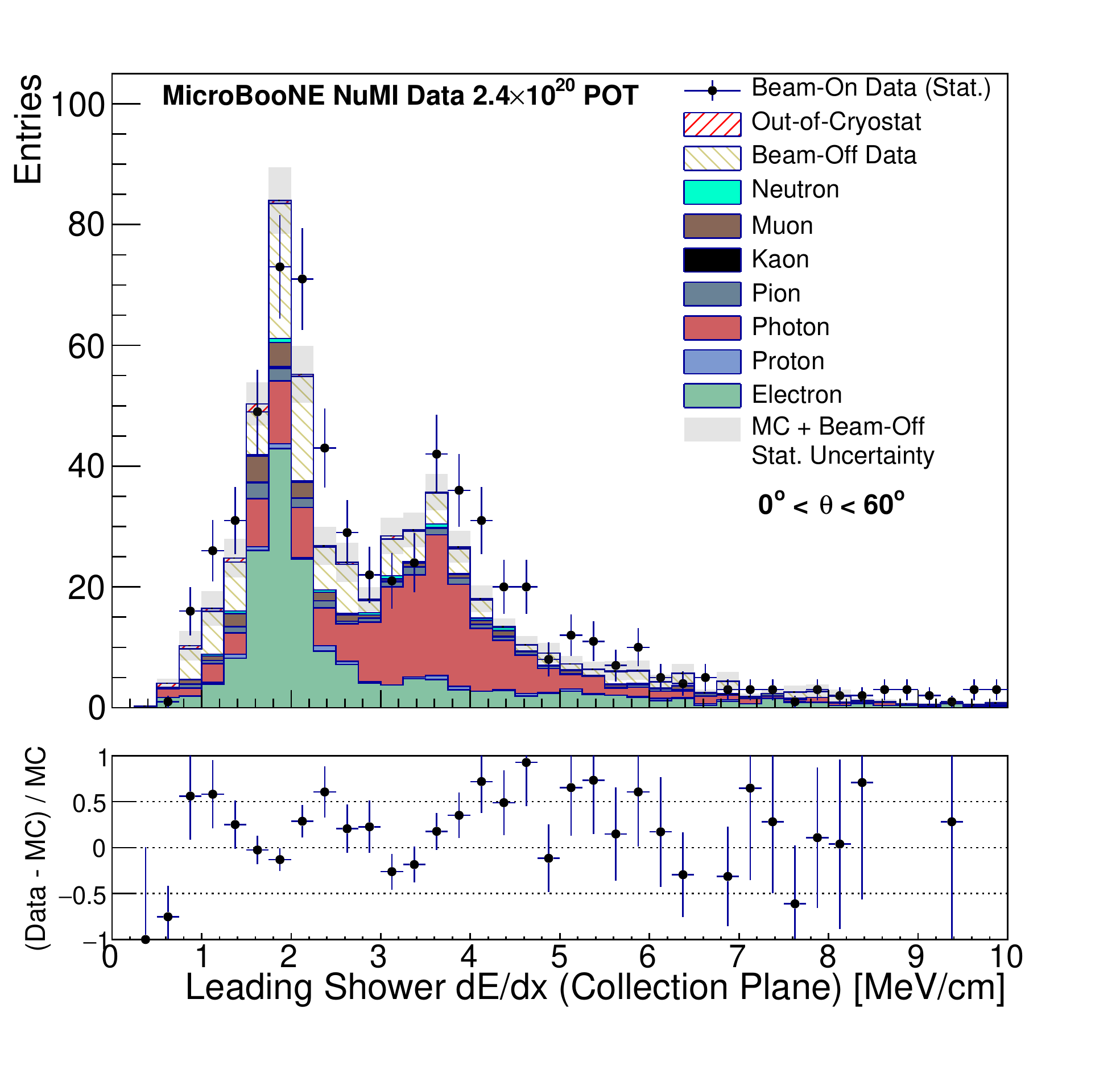}
\caption{dE/dx of leading showers for neutrino candidates broken down by particle type. This plot is made for leading shower $\theta$ between 0$^{\circ}$ and 60$^{\circ}$ where the reconstruction of showers is good. Electrons are gathered in the MIP peak, while most photons are around 4 MeV/cm. }
\label{post_cuts_dedx_true_particle}
\end{figure}

We find that the dE/dx variable is more effective in removing photon-induced backgrounds. Figure~\ref{post_cuts_dedx_true_particle} illustrates its separation power in rejecting the photon-like events which dominate around the 4~MeV/cm peak in the dE/dx distribution.


\subsection{Selection Performance}

Many of the selection requirements focus on removing cosmic ray interactions reconstructed as showers. The overall decrease in the cosmic ray contamination is a factor of $10^{5}$ compared to the initial \texttt{Pandora} reconstruction stage which can have many reconstructed cosmic rays in a readout window. This ultimately brings the cosmic ray contamination to roughly the same size as the number of selected electron neutrino and antineutrino interactions. The remaining non-cosmic backgrounds contribute to approximately 19\% of the selected neutrino candidates. This demonstrates the selection's ability to reliably remove beam-induced backgrounds such as $\nu_{\mu}$ CC and $\pi^{0}$ interactions.  We find that the selection is sensitive to neutrino events where the final state electron momentum is higher than 48 MeV/c, and includes the entire angular phase space of the electron.

\begin{figure}[htb]
\includegraphics[width=0.49\textwidth]{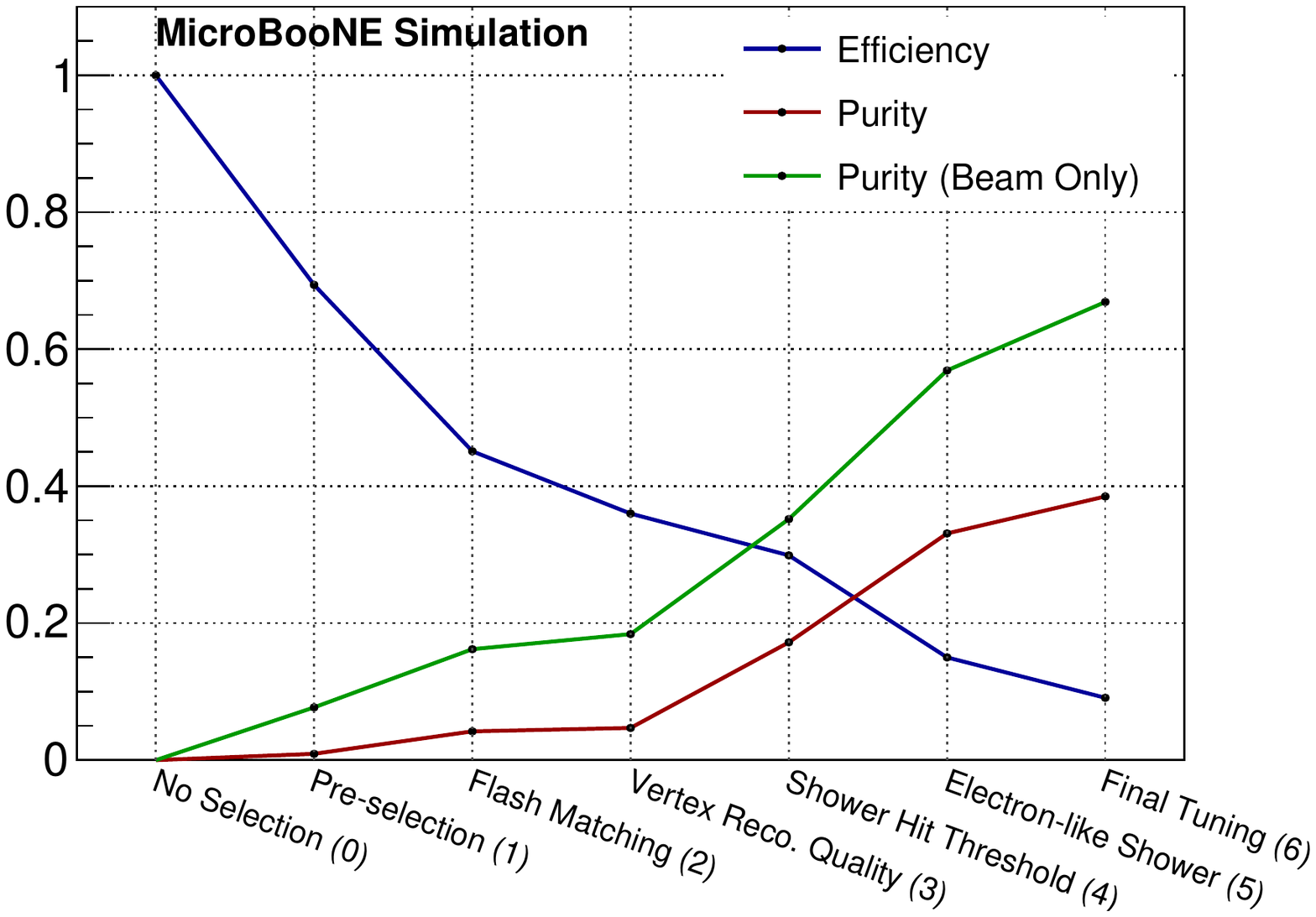}
\caption{Summary of the selection performance, including efficiency, purity, and purity (beam only). The beam only purity corresponds to the selection purity if cosmic backgrounds could be completely removed.}\label{efficiency_purity}
\end{figure}

A summary of the purity and efficiency at different stages of the selection can be seen in Fig.~\ref{efficiency_purity}. The steepest decrease in the efficiency, once the basic pre-selection is applied, occurs with the application of the shower opening angle and dE/dx selection requirements which are both included in the electron-like shower category. For the latter the decrease occurs because of the difficulty in calculating the dE/dx for the shower direction of certain signal events. This is also the step with the largest increase in purity. With improvements to calculating dE/dx for all shower directions, this variable is expected to become even more powerful. The beam-only purity (in green) is shown in Fig.~\ref{efficiency_purity} and has a final value above 60\%. This is calculated by considering only beam-induced backgrounds which would be the ideal case if all cosmic ray backgrounds in this analysis could be completely removed. In future analyses with improved cosmic rejection tools such as using the cosmic ray tagger system installed around MicroBooNE, contamination from cosmic ray backgrounds should significantly decrease which will enable much higher purities and improved selection performance. The final selection efficiency is $9.1\% \pm 0.3\%$ for the $\nu_e$ + $\bar{\nu}_e$ sample, which can be divided into $8.8\% \pm 0.4\%$ and  $12.2\% \pm 1.0\%$ for $\nu_{e}$ and $\bar{\nu}_e$ respectively.